\documentclass[12pt]{article}
\usepackage{amsmath,amssymb,amsfonts,graphicx,verbatim,cite}
\usepackage{wrapfig}
\pdfoutput=1

\usepackage[usenames,dvipsnames]{xcolor}

\usepackage{hyperref}
\hypersetup{colorlinks=true, linkcolor=BrickRed, citecolor=Violet, filecolor=OliveGreen, urlcolor=RoyalBlue, filebordercolor={.8 .8 1}, urlbordercolor={.8 .8 0}}

\definecolor{palatinatepurple}{rgb}{0.41, 0.16, 0.38}

\definecolor{uglybrown}{rgb}{0.8,  0.7,  0.5}
\definecolor{darkgreen}{rgb}{0,.5,0}

\def\lsim{ \lower .75ex \hbox{$\sim$} \llap{\raise .27ex
\hbox{$<$}} }
\def\gsim{ \lower .75ex \hbox{$\sim$} \llap{\raise .27ex
\hbox{$>$}} }

\renewcommand{\title}[1]{\vbox{\center\LARGE{#1}}\vspace{5mm}}
\renewcommand{\author}[1]{\vbox{\center#1}\vspace{5mm}}
\newcommand{\address}[1]{\vbox{\center\em#1}}

\renewcommand{\title}[1]{\vbox{\center\bf{\Large{#1}}}\vspace{5mm}}

\numberwithin{equation}{section}

\newcommand{\beq}{\begin{equation}}
\newcommand{\eeq}{\end{equation}}
\newcommand{\be}{\begin{equation}}
\newcommand{\ee}{\end{equation}}

\def\bea{\begin{eqnarray}}
\def\eea{\end{eqnarray}}

\def\({\left(}
\def\){\right)}

\def\CA{{\cal A}}

\def\CC{{\cal C}}

\def\CN{{\cal N}}
\def\CO{{\cal O}}

\def\CW{{\cal W}}

\newcommand{\ket}[1]{| #1 \rangle }

\newcommand{\TFD}{|\text{TFD}\rangle}
\newcommand{\lads}{\ell_{\text{AdS}}}

\renewcommand{\eqref}[1]{Eq.~\ref{#1}}

\begin{document}

\begin{titlepage}

\hfill \\
\hfill \\
\vskip 1cm

\title{Complexity, action, and black holes}

\author{Adam R. Brown${}^a$, Daniel A. Roberts${}^b$, Leonard Susskind${}^a$,  \\ Brian Swingle${}^a$, and Ying Zhao${}^a$ }

\address{${}^a$Stanford Institute for Theoretical Physics and\\
Department of Physics, Stanford University,\\
Stanford, California 94305, USA}

\address{${}^b$Center for Theoretical Physics and\\
Department of Physics, Massachusetts Institute of Technology,\\
Cambridge, Massachusetts 02139, USA}

\begin{abstract}

Our earlier paper ``Complexity Equals Action" conjectured that the quantum computational complexity of a holographic  state is given by the classical action of a region in the bulk (the ``Wheeler-DeWitt'' patch). We provide calculations for the results quoted  in that paper, explain how it fits into a broader (tensor) network of ideas,  and elaborate on the hypothesis that black holes are the fastest computers in nature.

\end{abstract}

\vfill

\end{titlepage}

\vfill\eject

\tableofcontents

\vfill\eject

\section{Introduction}

\begin{figure}
\label{fig:CVduality}
\centering
\includegraphics[width=.9\textwidth]{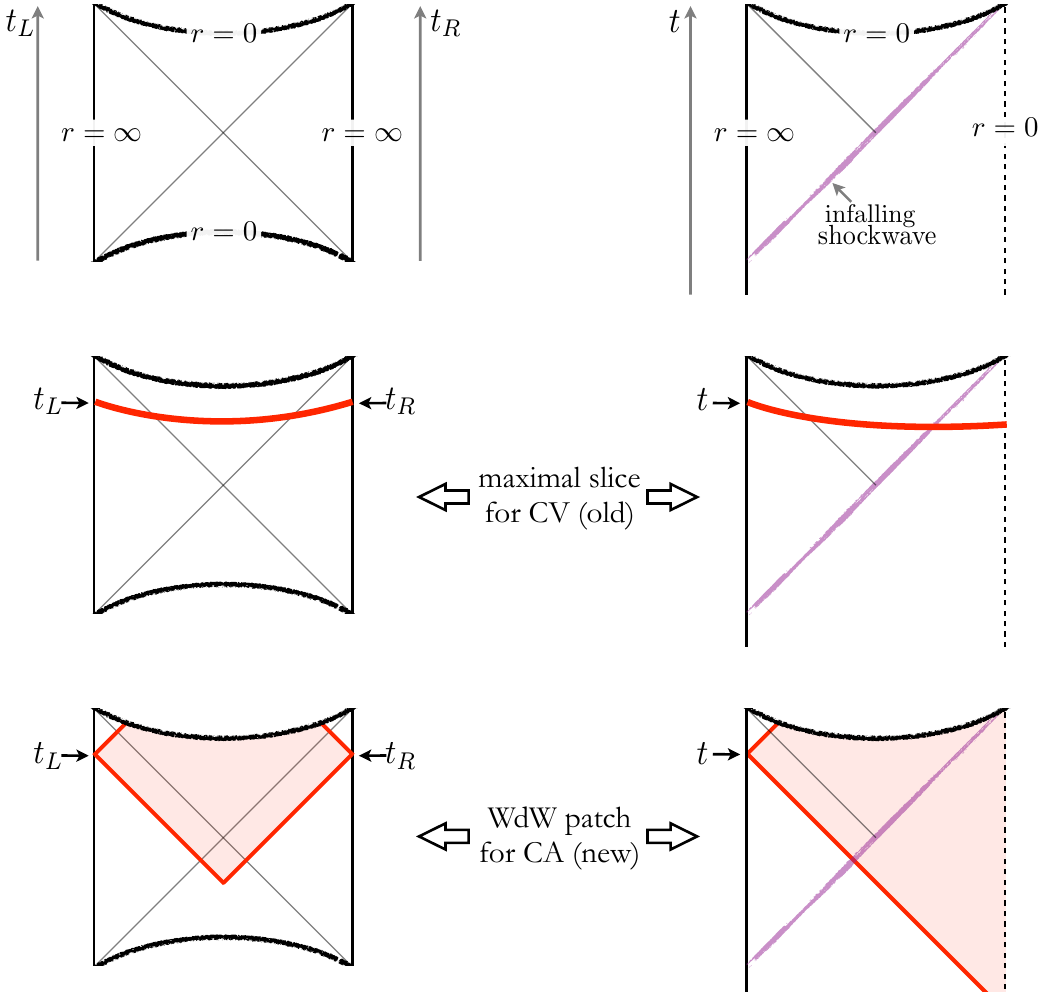}
\caption{The Penrose diagrams for two-sided eternal black holes (left) and one-sided black holes that form from collapsing shock waves (right). The two-sided black hole is dual to an entangled state of two CFTs that live on the left and right boundaries; the one-sided black hole is dual to a single CFT. The (old) complexity/volume conjecture related the complexity of the entangled CFT state to the volume of the maximal spatial slice anchored at the CFT state. Our (new) complexity/action conjecture relates the complexity of the CFT state to the action of the Wheeler-DeWitt patch.}
\end{figure}

According to the holographic principle all physics is encoded at the boundaries of spacetime.  A good deal is known in the context of AdS/CFT about this encoding as long as we restrict our attention to the region outside horizons. Very much less is known about the holographic encoding  of physics behind black hole horizons. The  purpose of this paper is to explore a proposal for how properties of the black hole interior are represented on the holographic boundary. We propose that the quantum complexity of the boundary state is equal to the classical action of a spacetime region that extends deep inside the horizon. This region---which we call the Wheeler-DeWitt patch---is defined as the bulk domain of dependence of a bulk Cauchy slice anchored at the boundary state. For the case of one- and two-sided black holes in Anti-de Sitter space, the geometry is shown in Fig.~\ref{fig:CVduality}. Quantum complexity $\CC$ is the minimum number of elementary operations (quantum gates) needed to produce the boundary state of interest from a reference state. Calling the action of the Wheeler-DeWitt patch $\CA$, our proposal is \cite{Brown:2015bva}
\be
\CC = \frac{\CA}{\pi \hbar}.
\label{eq:CAformula}
\ee

Quantum complexity entered black hole physics to help quantify the difficulty of decoding Hawking radiation \cite{haydenharlow}, but it appears to also shed light on physics behind the horizon. It was observed that black hole interiors grow in time long after local equilibrium is reached \cite{hartmanmaldacena}. The complexity of the plasma dual to the black hole is also expected to increase long after local thermal equilibrium is reached, so it was proposed that the two growths are dual descriptions of the same phenomenon  \cite{Cgeodesicduality,Susskind:2014rva}. In this paper we will characterize the size of the black hole interior by its action. This action conjecture subsumes a previous conjecture \cite{complexityshocks} that characterized the size of the black hole interior by its spatial volume.

The original complexity/volume  or ``CV'' duality \cite{complexityshocks} stated that the complexity of the boundary state is proportional to  the spatial volume $V$ of a maximal slice behind the horizon,
\beq
\CC \sim \frac{V}{G \ell},
\label{CV--duality}
\eeq
where  $\ell$ is a length scale that has to be chosen appropriately for the configuration (typically either the AdS radius or the radius of the black hole).

 CV-duality had a number of nice features. The maximal volume slice is a relatively robust geometric object that naturally grows at a rate roughly proportional to the temperature $T$ times the entropy $S$ of the black hole. CV-duality works because $TS$ provides a rough estimate for the rate of complexification of a strongly coupled conformal plasma. Shock wave geometries dual to the insertion of a simple operator in the distant past  provide additional tests of CV-duality. For these shock wave geometries it was found that the maximal volume slice had cancellations that precisely match cancellations expected in the complexity \cite{complexityshocks,localshocks, Susskind:2014jwa}. Finally, various tensor network constructions qualitatively agreed with the holographic results \cite{complexityshocks,localshocks}. (For additional work on the CV-duality, see also \cite{Alishahiha:2015rta,Barbon:2015ria,Barbon:2015soa}.)

However,  CV-duality has a number of unsatisfactory elements. First we must choose by hand a length scale $\ell$, depending on the situation. Further, it is also not clear why the maximal volume slice should play a preferred role. The conjecture that complexity equals action  (``CA-duality") inherits all the nice features of CV-duality and none of the unsatisfactory elements.

One virtue of CA-duality is that it associates with the boundary state the entire Wheeler-DeWitt patch, and doesn't single out a special spacelike slice. Another virtue is that it is not necessary to introduce an arbitrary length scale; as a consequence, the prefactor of the complexity could  potentially have physical meaning. CA-duality reproduces the results of CV-duality for small and large AdS black holes without the need to choose different lengths in the two cases. Once the constant in Eq.~\ref{eq:CAformula} has been fixed by considering any particular black hole, the conjecture then makes unambiguous predictions with no further free parameters for black holes of any size, for black holes in any number of dimensions,  for black holes that are charged or rotating, and for states that are not black holes. The CA-duality passes all the same shock wave tests as were passed by CV-duality.

We work in the context of AdS/CFT and assume a familiarity with those notions; we also assume familiarity with notions of quantum information including quantum circuits, see e.g. \cite{complexityreview1,complexityreview2}. CA-duality is a quantum-classical duality, insofar as it relates a highly quantum theory on the boundary (where the complexity is defined) to a highly classical theory in the bulk (where the action is defined). 
In order to use CA-duality we will need to identify a set of weakly interacting semiclassical bulk degrees of freedom, to consider states that are semiclassical with respect to those degrees of freedom, and  to assume the existence of a unique real-valued semiclassical effective action for those degrees of freedom. In this paper we will generally be considering semiclassical bulk states of the Einstein-Maxwell system.
%
When studying states with growing complexity we will always restrict to times less than $O(e^S)$. Beyond this time the complexity is expected to saturate and fluctuate (and eventually decrease at times of order the quantum recurrence time $O(e^{e^S})$).

The paper is structured as follows:

In \S\ref{subsec:CA} we more precisely define our conjecture.

In \S\ref{sec:infobound} we review bounds on information processing and conjecture that black holes are the fastest computers in nature.

In \S\ref{sec:bh} we examine a number of different types of black holes and compute the growth of the action of the Wheeler-DeWitt (WDW) patch with time.

In \S\ref{sec:testingwithbh} we use our conjecture to compare the rate of growth of action of black holes with a conjectured quantum mechanical bound on complexity growth inspired by a conjectured bound of Lloyd's \cite{lloyd2000ultimate}. We find precise agreement for uncharged and rotating black holes. For charged black holes, the situation appears more complicated, but these complications result from  dynamical issues associated with the production of hair, and we have no controlled examples that violate our bound.

In \S\ref{sec:staticshells} we test our conjectures with an analysis of black holes surrounded by static shells of inert matter. We find our conjectures pass.

In \S\ref{sec:shockwaves} we test our conjectures with an analysis of black holes perturbed by shock waves. We show that CA-duality passes this test, for much the same reason that CV-duality did. We find that the geometry of the Einstein-Rosen bridge of shock wave states matches the geometry of the minimal tensor network that describes such products of precursors. In general, the CA-calculations turn out to be significantly simpler than the CV calculations. We also discuss {complexity growth with} finite-energy {perturbations}.

In \S\ref{sec:tensornetworks} we will provide a tensor network computation of complexification. 
Using tensor network renormalization group techniques, we will show that the growth of complexity is independent of the UV cutoff.

In \S\ref{sec:discussion} we review the main implications of CA-duality presented in this work, comment on a number of extensions and puzzles raised by our work, and discuss some open questions.

\subsection{Complexity equals action}
\label{subsec:CA}

Here we define  CA-duality  within the context of holographic duality. The setup of CA-duality and the relationship between CV-duality and CA-duality is illustrated in Fig.~\ref{fig:CVduality}. Choosing times $t_L , t_R$ on the left and right AdS boundaries of
 the eternal AdS-Schwarzschild black hole determines a state $\ket{\psi(t_L,t_R)}$,
\beq
\ket{\psi(t_L,t_R)} = e^{-i (H_L t_L + H_R t_R)} \TFD,
\eeq
where $\TFD = Z^{-1/2} \sum_\alpha e^{-\frac{\beta E_\alpha}{2} } \ket{E_\alpha}_L\ket{E_\alpha}_R$ is the thermofield double (TFD) state which purifies the thermal state of one side.\footnote{Note that with our conventions, time on both sides of Fig.~\ref{fig:CVduality} increases upwards.}

CV-duality proposes that the complexity of $\ket{\psi(t_L,t_R)}$ is computed from the volume $V$ of a $(D-1)$-dimensional maximal volume slice anchored at $t_L$ and $t_R$,
\beq
\CC(\ket{\psi(t_L,t_R)})
 \sim
\frac{V(t_L,t_R)}{G \ell},
\label{old}
\eeq
where $G$ is Newton's constant and $\ell$ is another length scale (the AdS radius for large black holes and the Schwarzschild radius for small black holes). This yields a complexity which grows linearly with time and is proportional to temperature $T$ and entropy $S$, a result that roughly matches CFT expectations.

The prescription Eq.~\ref{old} has a degree of arbitrariness. First of all the choice of foliation by maximal slices is not  unique. Moreover these slices do not foliate the entire geometry behind the horizon. Another unattractive feature is the introduction of the length scale $\ell$ which varies from case to case. The proposal of this paper remedies
 these unsatisfactory elements while retaining all the benefits of CV-duality.

To motivate the CA proposal we note that the geometry of the $D$-dimensional world volume $\CW$ behind the horizon is roughly that of a $D$-dimensional tube of length $t_L+t_R.$  The (\emph{D}-1)-dimensional cross section of the tube has spatial area $GS$ and time duration $\sim \ell_{\text{AdS}}.$ The world volume of the tube is defined to be $|\CW|$.
 Since $V(t_L,t_R) \sim G S (t_L + t_R)$ the volume formula may be expressed as
\beq
\frac{V(t_L,t_R)}{G \ell_{\text{AdS}}} \sim \frac{|\CW(t_L,t_R)|}{G \lads^2}.
\eeq
Since the cosmological constant $\Lambda$ is proportional to $-1 / \lads^2$, the second expression is roughly the classical action of the world volume $\CW$. This clue leads to CA-duality which we now define.

Consider again the eternal black hole geometry (the extension to other geometries is straightforward) and pick times $t_L$ and $t_R$. Define the Wheeler-DeWitt (WDW) patch $\CW(t_L,t_R)$ to be the union of all spacelike surfaces anchored at $t_L$ and $t_R$. The WDW patch is equivalent to the spacetime region sandwiched between forward and backward light rays sent from the boundary at $t_L$ and $t_R$. $\CW$ is also the spacetime region determined by the data on the maximal slice (or any other spacelike slice within $\CW$) and is naturally associated with the boundary state, e.g. \cite{transhorizon}. Note that $\CW$ is not a causal patch so no single observer is able to monitor all of $\CW$; this is consistent with the complexity not being a conventional quantum mechanical observable.

Let $\CA_\CW$ be the action obtained by integrating the bulk Lagrangian over $\CW$ and including suitable boundary terms on $\partial \CW$ (see \S\ref{sec:bh}). The complexity $\CC$  (defined more fully just below) is roughly the minimum number of gates from a universal set necessary to produce the state of interest from a simple state. CA-duality states that
\beq
\boxed{
\CC(\ket{\psi(t_L,t_R)}) = \frac{\CA_\CW}{\pi \hbar}.} \label{eq:Boxed-CA}
\eeq
The factor of $\pi$ is arbitrarily chosen so that an increase of complexity by one unit advances the phase of $e^{i \CA_\CW/\hbar}$ from one to minus one.
At present we do not assign a physical meaning to this factor of $\pi$, it is only a convention to normalize the complexity.  We speculate in Appendix~\ref{sec:boundapp} that there is a more universal continuum version of complexity where the prefactor can be unambiguously defined.

\section{Bounds on information storage and processing}
\label{sec:infobound}

Given limited resources, there are limits on the storage and processing of information. For the purposes of our conjecture we are ultimately interested in information processing, but the bounds on information storage are so much better studied and understood that we will review them first as a model example.

\subsection{Bounds on information storage}

When the only scarce resource is spacetime, the information that may be stored is given by the holographic bound \cite{Hooft:1993gx,Susskind:1994vu}. The total information is bounded by the area,
\begin{equation}
S \leq \frac{\textrm{Area}}{4G \hbar} .
\end{equation}
When energy is also scarce, the total information is limited by the Bekenstein bound \cite{Bekenstein:1980jp, Marolf:2003sq, Casini:2008cr}
\begin{equation}
S \leq \frac{2 \pi E R}{\hbar},
\end{equation}
which is in general tighter. When only energy, and not space, is scarce, there is no limit on the information that may be stored: a box of thermal radiation has entropy
\begin{equation}
S \sim \left(\frac{E R}{\hbar}\right)^\frac{3}{4},
\end{equation}
which is arbitrarily large for an arbitrarily large box, $R \rightarrow \infty$.

Black holes saturate the holographic bound.\footnote{Modern hard-drives are very far from these limits. A 2015-era laptop SSD weighs 100 grams, has an area of 100$cm^2$ and stores 1TB =  $8 \times 10^{12}$ useful classical bits. This is a factor of $10^{28}$ below the $10^{41}$ qubits of the Bekenstein bound, and a factor of $10^{54}$ below the $10^{67}$ qubits of the holographic bound. Moore's ``law'' predicts we will run up against the fundamental limits at the start of the twenty-third century.} This is required by unitarity---if an object can be forced to undergo gravitational collapse by adding mass, then the second law of thermodynamics insists it must have less entropy than the resulting black hole. Black holes are the densest hard drives in nature.

\subsection{Bounds on information scrambling or chaos}
 For a system with $N$ degrees of freedom, the scrambling time $t_*$ is a measure of how long it takes for information about a small $O(1)$ perturbation to spread over $O(N)$ degrees of freedom \cite{fastscramble}. Another useful definition \cite{Page:1993df} is that this is the time required for reduced density matrix of any subsystem with fewer than half the degrees of freedom  to look approximately thermal.\footnote{In general, this concept is only useful for systems described by  $k$-local Hamiltonian with $k \ll N$. For more discussion, see Appendix~\ref{sec:appendix:A1}}

 Black holes are the fastest scramblers in nature \cite{bhmirror,fastscramble,Maldacena:2015waa}. There are two senses in which this is true:

 First, for a system arranged on a $d$-dimensional lattice, the scrambling time can be no faster than $t_* \sim N^{1/d}$, where $d$ is the number of spatial dimensions of the system \cite{fastscramble}. $N^{1/d}$ is proportional to the system's linear dimensions and implies that the  scrambling is ballistic. It was conjectured \cite{bhmirror,fastscramble} (see also \cite{Lashkari:2011yi}) and then shown \cite{shock} that the scrambling time for black holes in AdS is given by
 \be
t_* \sim \beta \log N.
 \ee
For black holes the dimensionality of the lattice $d$ is effectively infinite.

Second, not only are black holes the fastest scramblers because the scrambling time is logarithmic in $N$, they also have the smallest possible coefficient in front of the logarithm
 \be
t_* = \lambda_{L}^{-1} \log N,
 \ee
where the rate of scrambling can be interpreted in terms of chaos as a Lyapunov exponent $\lambda_L$ \cite{kitaev,Maldacena:2015waa}. A useful measure of the strength of chaos is given by the out-of-time-order four-point correlator \cite{LarkinOvchinnikov}
\be
\langle W(t)\, V \,W(t)\, V\rangle_\beta,
\ee
where $\langle \cdot \rangle_\beta$ is the thermal expectation value, $W$ and $V$ are simple local Hermitian operators, and $W(t) \equiv e^{iHt} W e^{-iHt}$. In chaotic systems, this correlation function will exhibit chaotic Lyapunov behavior (the butterfly effect) and initially decay as \cite{LarkinOvchinnikov, kitaev, Maldacena:2015waa}
\be
\langle W(t)\, V \,W(t)\, V\rangle_\beta = f_0 - f_1\, e^{\lambda_L (t - t_*)} + O(N^{-2}),
\ee
with $O(1)$ constants $f_0, f_1$.
Quantum mechanics puts a bound on this exponent \cite{Maldacena:2015waa}
\be
\lambda_L \le \frac{2\pi k_B T}{\hbar},
\ee
and black holes saturate this bound \cite{shock,Shenker:2013yza,localshocks,kitaev,Shenker:2014cwa}.

\subsection{Bounds on quantum evolution}

As well as being concerned with how much information may be stored, or how fast information may be scrambled, we are interested in how fast the system can change its global state. The time for the system to reach an orthogonal state is perhaps the simplest way to measure the rate of change of the global state. In this direction, there are two proved theorems, the Aharonov-Anandan-Bohm bound and the Margolus-Levitin bound. However, the orthogonality time is not the right measure to associate with the growth of the black hole interior, so we turn instead to the rate of growth of quantum gate complexity to measure how the global state is changing. The existing bounds on complexity growth are UV sensitive and not useful for our purposes, so we conjecture a new bound on the rate of growth of complexity inspired by Lloyd's conjecture. 

We first review the bounds on orthogonality time. Then we recall Lloyd's conjecture and formulate our own conjecture on the rate of complexification.

The Aharonov-Anandan-Bohm bound involves the standard deviation of the energy,
\begin{equation}
\textrm{orthogonality time} \ \geq \frac{\pi \hbar }{2  \Delta E} .
\end{equation}
Since time is not an operator in quantum mechanics, the energy-time uncertainty relation does not say that the time must be uncertain \cite{Anandan:1990fq,Aharonov:1961mka}, instead it says that time must be spent to evolve to an orthogonal state. The Aharonov-Anandan-Bohm bound is saturated by a two state system
\begin{equation}
| \psi (t) \rangle = \frac{1}{\sqrt{2}} \left( | 0 \rangle + e^{i E t} | E \rangle \right) \ \rightarrow  \ |\langle \psi(t) | \psi (0) \rangle | = \cos \frac{E t}{2 \hbar}. \label{eq:Heisenbergexample}
\end{equation}

The Margolus-Levitin bound involves the expectation value of the energy above the ground state \cite{Margolus:1997ih}
\begin{equation}
\textrm{orthogonality time} \geq \frac{\pi \hbar}{2 \langle E \rangle} .
\end{equation}
This is saturated by the same two state system that saturates the Aharonov-Anandan-Bohm bound, Eq.~\ref{eq:Heisenbergexample}. Depending on state, either of the Margolus-Levitin and Aharonov-Anandan-Bohm bounds may be tighter.

Inspired by these bounds on orthogonality time, Lloyd conjectured that they imply a bound on the rate of computation, loosely defined \cite{lloyd2000ultimate}. If we had a computer where after each time step the logical state of the computer was given by a classical bit string (a quantum state in the ``computational basis"), then because any two distinct bit strings correspond to orthogonal states the maximum rate at which the system can cycle through bit strings is given the tighter of the two orthogonality time bounds. Lloyd also noted that $N$ parallel copies of a computer could be understood to compute $N$ times as fast; while the expected energy scales as $N$, the standard deviation of the energy would only increase by $\sqrt{N}$. In light of this, Lloyd conjectured \cite{lloyd2000ultimate} that by allowing parallel computation the total rate of computation is proportional to the average energy. This work is related to other bounds on computation and communication \cite{otherbounds1,otherbounds2,otherbounds3} and to some ideas about quantum gravity \cite{lloyd_qg1,lloyd_qg2,lloyd_qg3}.

To generalize and make precise Lloyd's notion of ``operations per second", we would like to consider how complexity builds up in an isolated unitarily evolving quantum system in a general quantum state. Building on the work of Aharonov-Anandan-Bohm, Margolus-Levitin, and Lloyd, we conjecture a similar bound on the rate of growth of complexity. Informally this is the rate of growth of the number of simple gates needed to prepare the state of the computer from a reference state,
\begin{equation}
\frac{d \textrm{gates}}{dt} \leq \frac{2E}{\pi \hbar}. \label{eq:LloydBound}
\end{equation}

\eqref{eq:LloydBound} is fast. A 100$g$ 2015-era CPU manages about $10^{12}$ classical operations per second,\footnote{The world's fastest ``supercomputer'' is the collection of all the bitcoin miners, which have a total FLOP rate of 5 million petaFLOPs, about a thousand times more than all the single site supercomputers combined.} much less than the permitted $10^{50}$. The slowdown is largely the fault of wastefully locking up most of the energy in the rest mass of atoms, where it does no useful computation. (Indeed, counting just the energy in the logical qubits, modern cold atom experiments come close to saturating Eq.~\ref{eq:LloydBound}). A fast computer should avoid massive particles, and should interact strongly. A good candidate for a fast computer is therefore a highly energetic strongly coupled conformal field theory---these are known under some circumstances to be holographically dual to black holes.

Unlike in the case of information storage, there is no tight argument that black holes must be the fastest information processors in nature. There is no thought experiment that tells us that if a mountain of CPUs undergoes gravitational collapse, it must thereafter compute faster. Nevertheless, in light of the above considerations, it seems reasonable to conjecture that black holes excel as much here as they do in other information theoretic tasks. We conjecture that black holes are the fastest computers, in the sense that they saturate the complexification bound Eq.~\ref{eq:LloydBound}.\footnote{We have argued that black holes may be the fastest computers in nature, at least within some restricted class. Even assuming this is true, it is important to appreciate what this does and does not mean. It means that nothing can implement quantum gates any faster than a black hole (at fixed energy). It certainly means that nothing can simulate the complete output from a black hole any faster than the black hole itself can produce that output (at fixed energy). It does not mean that a black hole can beat you at chess.

This is not a phenomenon unique to black holes. For our purposes, a ``computer'' is a device that implements a unitary operation. This is useful if you are interested in implementing that unitary---if you are interested in the state of the Hawking radiation that emerges after a certain input has been thrown into the black hole, for example. If instead you have another question in mind---factorization, chess strategy, etc.---then the unitary is useful to you only if combined with other computing resources to ``encode'' the problem instance in terms of ingoing perturbations into the black hole, and to ``decode'' the Hawking radiation output of the black hole in terms of a strategy recommendation. For a generic problem, the computational resources needed to encode and decode the question are expected to be no smaller than the computational resources required to just solve the problem directly, without reference to the black hole. For generic problems, therefore, black holes don't help. (See Chapter~6 of \cite{DBLP:journals/corr/abs-1108-1791}.) \label{eq:footnotecomputionalism}}

\subsection{Complexity}

Given a state $\ket{\psi}$ on $n$ qubits, the state complexity of $\ket{\psi}$ is the minimum number of gates necessary to construct $\ket{\psi}$. We fix a simple reference state $\ket{\psi_0}$ and a universal gate set of elementary unitaries $\{G_\alpha\}$ and define the complexity of $\ket{\psi}$ to be the minimum number of gates necessary to produce $\ket{\psi}$ from $\ket{\psi_0}$. More generally, we require only that $\ket{\psi}$ be closely approximated by a state produced from $\ket{\psi_0}$. The complexity of $\ket{\psi}$ in principle depends on all the details of the construction, but is expected to be relatively robust in its basic properties, e.g. the distinction between states of $\text{poly}(n)$ complexity and $\text{exp}(n)$ complexity is independent of the details.

In this paper we conjecture the existence of a more refined notion of complexity appropriate for continuum field theories. Our refined notion of complexity should have the key property that, until it saturates, it grows linearly in time under evolution by a local Hamiltonian. The rate of complexity growth should also be proportional to the number of active degrees of freedom. We assume that a definition including all these details exists, at least for semiclassical bulk states, but many of our conclusions are robust even if such a unique definition cannot be constructed. An incomplete discussion of these open questions is given in Appendix~\ref{sec:boundapp}.

So long as we are considering the late time rate of change of complexity growth, the reference state $\ket{\psi_0}$ may be taken to be the thermofield double state. (For one-sided systems the reference state would be a locally perturbed ground state.) We will discuss the reference state in more detail in \cite{ToAppear}.

\subsection{Conjectured complexity growth bound}

We conjecture that the complexity can be defined such that the growth of complexity is bounded by the energy,
\beq
\label{eq:bound}
\frac{d}{dt} \CC( e^{-i H t}\ket{\psi}) \leq \frac{2 E_\psi}{\pi \hbar},
\eeq
where $E_\psi$ is the average energy of $\ket{\psi}$ relative to the ground state. To be precise, we conjecture that this bound holds for suitable semiclassical bulk states and similar field theory states; see Appendix~\ref{sec:boundapp}.

In calculations with black holes, $E_\psi$ will be the mass $M$ of the black hole. Hence we have defined the complexity so that black holes appear to saturate the Lloyd bound. We will give some evidence that other states of quantum gravity complexify more slowly. The complexity growth bound as we use it below for uncharged black holes thus reads
\beq
\boxed{\frac{d{\CC}}{dt} \leq  \frac{2M}{\pi\hbar}.}
\label{C-bound}
\eeq

All of our examples---uncharged black holes, rotating black holes, charged black holes, black holes surrounded by shells, and black holes perturbed by shock waves---obey this bound. Uncharged black holes saturate the bound.

\subsection{Bound on complexity growth with a conserved charge}\label{sec:conserved-charge}

The bound Eq.~\ref{C-bound} is general, but if the system carries conserved quantum numbers such as charge $Q$ or angular momentum $J$, a tighter bound might be possible. Intuitively this is because conserved charges provide a barrier to rapid complexification since some energy is tied up in noncomputing degrees of freedom. We first motivate the bound using the charged TFD state and then present the bound.

To define the charged version of the TFD state we introduce a chemical potential $\mu$. We may think of the chemical potential as an electrostatic potential that is positive on one side of the black hole and negative on the other. The TFD state is given by
\beq
| \textrm{TFD}_{\mu} \rangle = Z^{-1/2} \sum_\alpha e^{-{\beta (E_\alpha +\mu Q_\alpha     )}/{2} } \ket{E_\alpha, Q_\alpha }_L\ket{E_\alpha, -Q_\alpha}_R
\label{TFDQ}
\eeq
The time evolution of the state is modified by the chemical potential,
\beq
\ket{\psi(t_L,t_R)} = e^{-i (H_L + \mu Q_L)t_L }  e^{-i (H_R - \mu Q_R)t_R }
| \textrm{TFD}_{\mu} \rangle .
\eeq
(Here $H_L$ and $H_R$ are the $\mu = 0$ Hamiltonians.)
By the same argument that lead to the $\mu =0$ bound, the complexification bound becomes
\beq
\boxed{\frac{d\CC}{dt} \leq  \frac{2}{\pi \hbar} \left[(M - \mu Q) - (M - \mu Q)_{\text{gs}}\right]} \,\,\, \text{(charged black holes),}
\label{CQ-bound}
\eeq
where the subscript $gs$ indicates the state of lowest $(M-\mu Q)$ for a given chemical potential $\mu$.\footnote{$M-\mu Q$ is natural when viewing $\mu$ as an electrostatic potential which modifies the total energy. For another perspective on the proposed bound, note that the complexity of $e^{i  \mu Q t}$ oscillates in time with period $2\pi/\mu$ because $Q$ has integer spectrum. We believe that the combination $M - \mu Q$ serves to remove this rapidly oscillating portion of the complexity.} For rotating black holes the bound becomes
\beq
\boxed{\frac{d\CC}{dt} \leq  \frac{2}{\pi \hbar} \left[(M - \Omega J) - (M - \Omega J)_{\text{gs}}\right]}  \,\,\, \text{(rotating black holes)},
\label{CR-bound}
\eeq
where the role of the chemical potential is played by the angular velocity
$\Omega$.

Bounds of the type Eq.~\ref{CQ-bound} and Eq.~\ref{CR-bound} can be understood using the thermodynamic relation,
\be
d(M-\mu Q) = TdS
\label{thermo-relation}
\ee
or its integrated form,
\be
\left[(M - \mu Q) - (M - \mu Q)_{\text{gs}}\right]= \int_\textrm{gs}^S T dS.
\label{integrated-thermo}
\ee
In other words $\left[(M - \mu Q) - (M - \mu Q)_{\text{gs}}\right]$ is the heat content or internal energy relative to the ground state at fixed $\mu.$  One way to state our hypothesis that black holes are the fastest computers is that they efficiently use all of their internal energy to complexify.\footnote{In earlier work the integral in Eq.~\ref{integrated-thermo} was replaced by the simpler expression $TS.$ Generally the two are approximately the same, but in \S\ref{sec:testingwithbh} we will encounter a case in which they are not. This is the case of near-extremal large charged AdS black holes.} In other words black holes saturate the bound
\be
\frac{d\CC}{dt} \leq \int_\textrm{gs}^S T dS.
\ee

For rotating black holes in $D=3$ we find that the bound is saturated. For charged black holes the situation is more complicated. Assuming the true ground state is empty AdS or the extremal black hole at fixed chemical potential as appropriate, we can show that the bound is sometimes saturated but generally \textit{violated}. However, proper evaluation of the bound raises a difficult dynamical question, specifically the nature of the ground state at nonzero chemical potential. The nature of the ground state depends on the full operator content of the conformal field theory and hence requires a proper UV completion of Einstein-Maxwell theory. At least we must specify the light charged degrees of freedom.

In some cases supersymmetry provides additional constraints and the bound seems to be saturated. More generally the bound appears to be either qualitatively valid, up to an order one factor, or badly violated but in a situation where the true ground state is probably drastically different from the extremal black hole. We find the uncharged bound of Eq.~\ref{C-bound} is always obeyed.

\section{Action growth of black holes}
\label{sec:bh}

In this section, we will calculate the rate of increase of action of a Wheeler-DeWitt patch of the two-sided black hole (which we are conjecturing to be dual to the rate of growth of complexity of the boundary state).

We will consider the Einstein-Maxwell theory. The action is (using the conventions of \cite{Poisson})
 \begin{equation}
\CA =   \frac{1}{16 \pi G} \int_{\mathcal{M}}  \sqrt{|g|} \left( \mathcal{R} - 2 \Lambda  \right) - \frac{1}{16 \pi } \int_{\mathcal{M}}  \sqrt{|g|} F_{\mu \nu} F^{\mu \nu}  +  \frac{1}{8 \pi G} \int_{\partial \mathcal{M}} \sqrt{|h|} {K} , \label{eq:EinsteinHilbertActionForRealz}
\end{equation}
where the three terms are the Einstein-Hilbert (EH) action including a negative cosmological constant, the Maxwell electromagnetic action, and a York, Gibbons, Hawking (YGH) surface action.  In defining the extrinsic curvature $K$
 spacelike normals are taken to point outwards and timelike normals  inwards.

 A Reissner-Nordstrom-AdS black hole has metric
 \begin{eqnarray}
 ds^2 & = & - f(r)dt^2 + \frac{dr^2}{f(r)} + r^2 d \Omega_{D-2}^2 \\
f(r)  & = &   1 - \frac{8 \pi}{(D-2)\Omega_{D-2}} \frac{2 G M}{r^{D-3}} + \frac{8 \pi}{(D-2)\Omega_{D-2}} \frac{G Q^2 }{r^{2(D-3)}}+ \frac{r^2}{\lads^2} .
 \end{eqnarray}
Horizons occur where $f(r) =0.$ The radial coordinate at the (unique) horizon of a neutral black hole will be labeled
$r_h.$ The outer and inner horizons of the charged black hole are at $r_+$ and $r_-.$

 \subsection{Action of an uncharged black hole} \label{sec:actionuncharged}

\begin{figure}[h] 
   \centering
   \includegraphics[width=5in]{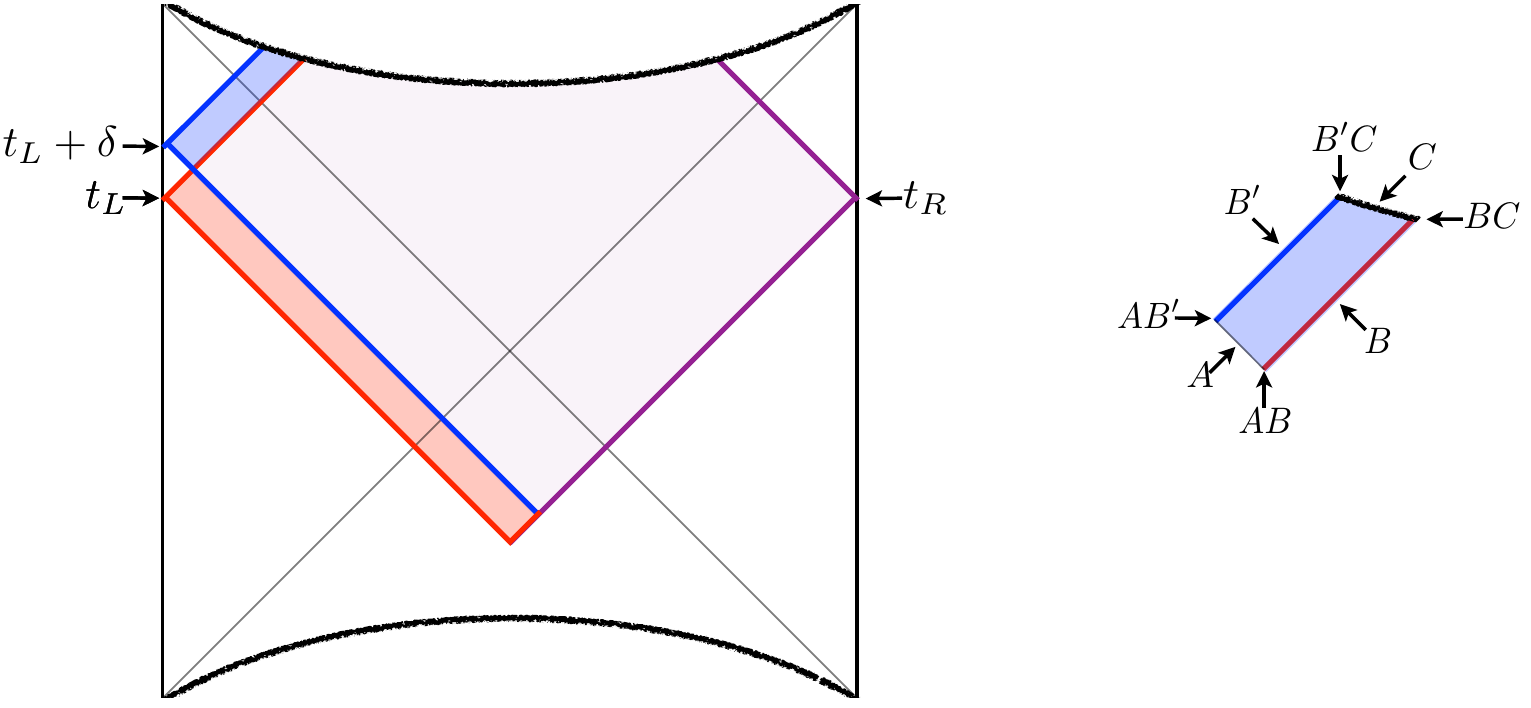}
   \caption{An uncharged AdS black hole. When $t_L$ increases, the Wheeler-DeWitt patch gains a slice (in blue) and loses a slice (in red).}
   \label{fig-incrementaluncharged}
\end{figure}

 Fig.~\ref{fig-incrementaluncharged} shows the Wheeler-DeWitt patch bounded by $t_L$ on the left and $t_R$ on the right for an uncharged ($Q=0$) black hole. As time passes on the left boundary, the patch grows in some places (shown in blue) and shrinks in others (shown in red).

The total volume (and total action) of the patch outside the horizon is infinite but independent of time, due to the time-translation symmetry outside the hole.

The region behind the past horizon contributes at early times, but at late times ($t_L + t_R \gg \beta$) shrinks exponentially to zero. At late times, the size of the $D-2$-sphere is constant up to terms that are exponentially small in time. Consequently, we can use the two-dimensional Gauss-Bonnet theorem \cite{GaussBonnet} on the remaining two dimensions to say that the contribution of the wedge behind the past horizon is topological and must be independent of time at late times.\footnote{It should be pointed out that while the rate of change of action from the region behind the past horizon is exponentially small in time, that exponential could in principle be multiplied by a UV-divergent quantity. We address this unresolved subtlety in \S\ref{sec:corner} and in \cite{ToAppear}. Note that  it is not always the  case that an arbitrarily small region has arbitrarily small action. A simple illustration of this is provided by a two-sphere, whose action is topological and finite even as the sphere becomes arbitrarily small. We thank Henry Maxfield for discussion.}  At late times, the whole contribution comes from the part of the blue slice that lives inside the future horizon.

This region is shown on the right hand side of Fig.~\ref{fig-incrementaluncharged}. The boundary contribution at the light sheet $B'$ replaces the old contribution at $B$; since $B'$ and $B$ are related by the time-translation symmetry, this change does not affect the total action. Similarly the added corner contribution from $AB'$ cancels the removed corner contribution from $AB$,  and $B'C$ cancels $BC$. This leaves the surface contributions from $A$ (at $r=r_h$) and from $C$ (at $r=0$), as well as the bulk contribution.

Using the expression for the action in Eq.~\ref{eq:EinsteinHilbertActionForRealz}, and using Einstein's equation $\mathcal{R}_{\mu \nu} - \frac{1}{2} \mathcal{R} g_{\mu \nu} + \Lambda g_{\mu \nu} = 8 \pi G T_{\mu \nu}$ implies that
\begin{equation}
\Lambda = -\frac{(D-1)(D-2)}{2 \lads^2} \ \  \textrm{ and } \ \  \mathcal{R} = \frac{2D}{D-2} \Lambda, \label{eq:LambdaAndR}
 \end{equation}
so the EH contribution to the action is proportional to the spacetime volume
\begin{equation}
\frac{d\CA_\textrm{EH}}{dt_L}  
= -\frac{\Omega_{D-2} r_h^{D-1}}{8 \pi G \lads^2} . \label{eq:bulkunchargedcausalpatch}
 \end{equation}
The integral extends right down to the singularity, but receives only a small contribution from the immediate vicinity of $r=0$.

To calculate the YGH surface term we will use that the trace of the extrinsic curvature of a constant-$r$  surface is
\begin{equation}
K =  \frac{1}{2} n^r  \frac{\partial_r \left(   r^{2(D-2)  } f \right)  }{ r^{2(D-2) }f},
\label{eq:K}
\end{equation}
so the YGH contributions to the action at $A$ (at $r = r_h$)
and  $C$ (at $r=0$) are
\begin{equation}
\frac{d\CA_{\partial \mathcal{M}}}{dt_L} =  \left[  - \bigg( \frac{D-1 }{D-2}\bigg) M   + \frac{\Omega_{D-2} r^{D-3} }{8 \pi G}  \left( (D - 2 ) + (D-1) \frac{r^2}{\lads^2} \right) \right]^{r_h}_0 . \label{eq:rsunchargedboundary}
 \end{equation}

 The individual rates of change Eq.~\ref{eq:bulkunchargedcausalpatch} and Eq.~\ref{eq:rsunchargedboundary} are simple expressions in terms of the black hole mass, but combining them (using that $f(r_h) = 0$) we find the remarkably simple result,
  \begin{equation}
\boxed{\frac{d \CA}{dt_L} = 2M.} \label{eq:unchargedactionincrement}
 \end{equation}

Note that this result applies to all uncharged nonrotating  AdS black holes in any number of dimensions, whether small, intermediate, or large compared to the AdS length.\footnote{We have performed a completely classical calculation, with $\hbar = 0$. Quantum mechanics can destabilize sufficiently small black holes. In the canonical ensemble, large and intermediate black holes ($r_h \, \gsim \, \lads$) are stable. In the microcanonical ensemble, even much smaller black holes may be stable.}

\subsubsection*{Ambiguities avoided}
\begin{figure}
\centering
\includegraphics[width=.45\textwidth]{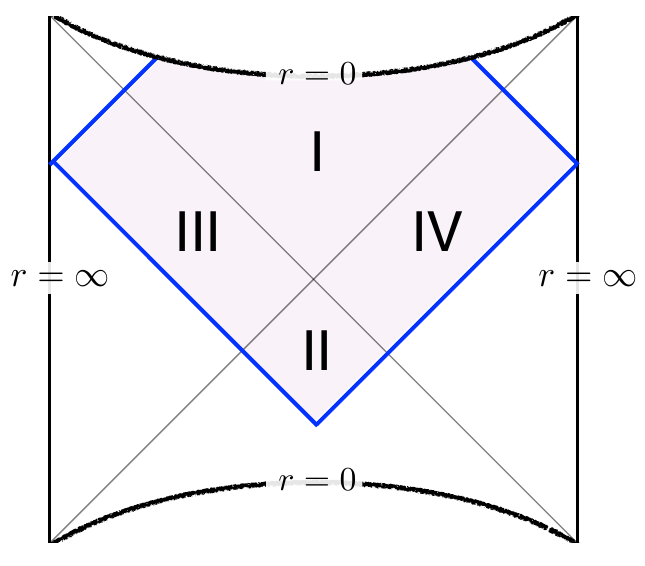}
\caption{The WDW patch divided into quadrants. In quadrants III and IV the WDW patch intersects the AdS boundary and causes a divergence in the action.}
\label{divergences}
\end{figure}

The action of a WDW patch is infinite because of the usual divergences near the AdS boundary. One also expects a matching divergence in the complexity  because of the large number of  UV degrees of freedom in the CFT dual. These divergences can cause ambiguities and nonuniversal behavior. Fortunately they do not contribute to the rates we have calculated. For example, in Fig.~\ref{divergences}  we have divided the WDW patch into four quadrants.
Two of the quadrants, III and IV,  reach the AdS boundary, and one can expect the action of these subregions to diverge.  The boost symmetry of the TFD state means that the contributions from III and IV are time independent and therefore do not contribute to the rate of change of complexity.

We mention this because a more ambitious calculation may attempt to estimate the complexity of formation of the  TFD state. We hope to come back to this \cite{ToAppear}, but in this paper we focus only on rates of change.

\subsection{Action of a charged black hole}
Adding electrical charge to an AdS black hole changes how the Wheeler-DeWitt patch terminates. Rather than terminating when the ingoing lights heets run into the singularity at $r=0$, it now terminates when the light sheets run into each other at $t=0$ (for $t_L = t_R$). Nevertheless, we can use much the same reasoning as before to find the late-time growth of the enclosed  action.

The entire WDW patch lies outside of the inner horizon at $r_-$. This is reassuring because it means the action is not sensitive to quantum instabilities of the inner horizon, so long as the horizon remains null. (We will not consider classical instabilities, which can lead to large changes in the structure of the inner horizon.)

\begin{figure}[h] 
   \centering
   \includegraphics[width=6in]{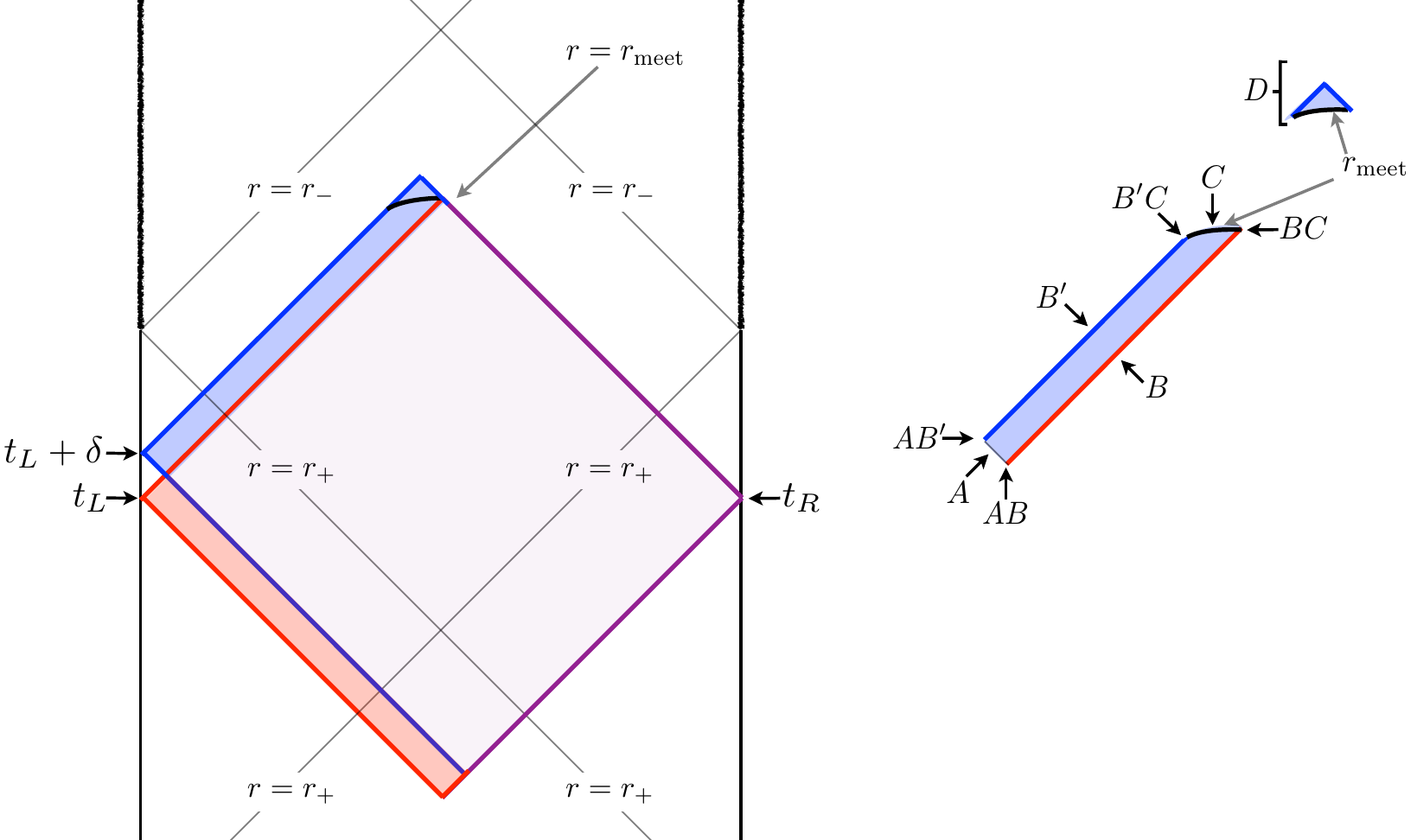}
   \caption{A charged AdS black hole. When $t_L$ increases, the Wheeler-DeWitt patch gains a slice (in blue) and loses a slice (in red). It is useful to consider separately the pieces above and below $r = r_\textrm{meet}(t_L,t_R)$. }
   \label{fig-incrementalcharged}
\end{figure}

As with the uncharged case, the action of that part of the patch that lies outside the outer horizon is independent of time, and the action of that part of the patch that lies behind the past horizon shrinks exponentially to zero at late times. The late-time rate of change of action comes from the part of the patch behind the future horizon, which is shown in Fig.~\ref{fig-incrementalcharged}.

The contribution from $B$ will cancel that from $B'$, the corners at $AB$ and $AB'$ will cancel, and the corners at $BC$ and $B'C$ will cancel. The contribution from $D$ is of order $\delta^2$ (since the size of the two-sphere is stationary at $D$, we are able to use the two-dimensional Gauss-Bonnet theorem there), and therefore does not contribute.  This leaves $A, C$, and the bulk term.

For simplicity we will work in $(3+1)$ dimensions where the  electromagnetic energy-momentum tensor is traceless. The electric field strength is
\begin{equation}
F_{r t} = - F_{t r}  = \frac{Q}{r^2} ,
\end{equation}
so  the Einstein-Hilbert-Maxwell (EHM) contribution to the on-shell action (Eq.~\ref{eq:EinsteinHilbertActionForRealz}) is
\begin{equation}
\frac{\CA_\textrm{EHM}}{dt_L}
=  \frac{r_+ - r_-}{2} \left( \frac{Q^2 }{r_+ r_-} - \frac{r_-^2 + r_+ r_- + r_+^2}{G \lads^2} \right) \label{eq:bulkchargedaction} .
\end{equation}
\noindent The York-Gibbons-Hawking surface actions at $A$ (at $r=r_+$) and at $C$ (at $r=r_-$) are
\begin{equation}
\frac{d\CA_{\partial \mathcal{M}}}{dt_L} =    \left[ -  \frac{3 M}{2} +  \frac{Q^2 }{2r}   + \frac{r }{G}  + \frac{3}{2G}  \frac{r^3}{\lads^2} \right]^{r_+}_{r-} . \label{eq:chargedAboundary}
 \end{equation}
 In total, the rate of change of action is (using that $f(r_+) = f(r_-) = 0$)
 \begin{equation}
\boxed{\frac{d\CA}{d t_L } =  \frac{r_+ - r_-}{G} \left( 1 + \frac{ r_-^2 + r_-  r_+ + r_+^2}{\lads^2 }   \right)  = \frac{Q^2 }{r_-} - \frac{Q^2 }{r_+}    . } \label{eq:chargedactionincrement}
 \end{equation}
Reassuringly, the total action reduces to the uncharged result Eq.~\ref{eq:unchargedactionincrement} when $Q \rightarrow 0$ (though how this total breaks up into bulk and boundary contributions differs).

 \subsubsection*{Near-extremal black holes}

 As extremality is approached, the complexification rate slows to a halt. The precise coefficient depends on the size of the black hole compared to the AdS length.

For very small  $(3+1)$-dimensional  charged black holes, extremality occurs at $M_\textrm{Q} = Q/\sqrt{G}$. The rate of change of action, whether near or far from extremality, is
\begin{equation}
\boxed{\frac{d \CA}{dt_L} \bigg|_{r_+ \ll \lads}  = 2 \sqrt{M^2 - Q^2/G} .}
\label{small-action}
\end{equation}

For very large  $(3+1)$-dimensional  charged black holes, extremality occurs at {$M_Q =\frac{2}{3} \left(\frac{G}{3} \right)^{-\frac{1}{4}  }\sqrt{ \frac{Q^3}{\lads}}$}. Near extremality
\begin{equation}
\boxed{\frac{d \CA}{dt_L}  \bigg|_{r_+ \gg \lads} = \sqrt{6} \sqrt{M_Q (M - M_Q)} \left( 1+ O\left( \frac{M - M_Q}{M_Q} \right) \right) .}
\label{3.15}
\end{equation}

\subsection{Action of a rotating BTZ black hole}\label{subsec-rotating}

A rotating black hole in $(2+1)$-dimensional AdS space has metric
\begin{equation}
ds^2=-f(r)dt^2+\frac{1}{f(r)}dr^2+r^2\left(d\phi-\frac{8GJ}{2r^2}dt\right)^2
\end{equation}
\begin{align}
f(r)=\frac{r^2}{\lads^2}-8G M+\frac{(8G J)^2}{4r^2} =\frac{(r^2-r_+^2)(r^2-r_-^2)}{r^2 \lads^2}.
\end{align}
The inner horizon ($r_-$) and outer horizon ($r_+$) are at
\begin{align}
&8G M=\frac{r_+^2+r_-^2}{\lads^2},\ \ \frac{8G J}{\lads}=\frac{2r_+r_-}{\lads^2}. \label{eq:BTZformulaforMJ}
\end{align}
The angular momentum is bounded above
\begin{equation}
M\geq\frac{J}{\lads}.
\end{equation}
This bound is saturated at extremality. The Penrose diagram for the rotating case is similar to that for the charged case in Fig.~\ref{fig-incrementalcharged}, and at late time, the EH contribution to the action of the Wheeler-DeWitt patch is
\begin{align}
\frac{d \CA_{\textrm{EH}}}{dt_L} 
 =-\frac{(r_+^2-r_-^2)}{4G \lads^2}.
\end{align}
The YGH surface contribution to the action can also be calculated, and as before it is $-2$ times the EH part, so the total rate of change of the action is
\begin{align}
\boxed{\frac{d\CA\textrm{ction}}{dt}=\frac{r_+^2-r_-^2}{4G \lads^2}=2\sqrt{M^2-\frac{J^2}{\lads^2}}.}
\label{rothole-action}
\end{align}
{Extremal rotating BTZ black holes ($J \rightarrow \lads M$) are all torque and no action.}

\section{Testing our conjectures with black holes} \label{sec:testingwithbh}

In this section, we will test our conjectures using the black holes considered in  \S\ref{sec:bh}.
The weak form is to check whether, assuming complexity equals action (Eq.~\ref{eq:Boxed-CA}), the rate of complexification of black holes satisfies the proposed bound on the rate of complexification (Eq.~\ref{C-bound}). The strong form will be to check whether the bound is saturated by black holes.

For neutral black holes we will find success; for all those that we have studied the complexification bound is exactly saturated. This includes static black holes of all masses in all dimensions as well as rotating BTZ black holes of any angular momentum.

Charged black holes are  murkier. We will find that small charged black holes saturate our bound, but that large charged black holes naively violate them. However, will find that in precisely those cases where our bounds are naively violated we have reasons not to trust the naive analysis.

\subsection{Neutral static black holes}

For neutral black holes, we calculated the rate of change of action of the WDW patch as
\begin{equation}
\frac{d \CA }{dt} = 2 M.
\end{equation}
The simplicity of this result underlies the claim that all nonrotating uncharged black holes
 saturate the bound, Eq.~\ref{eq:LloydBound}, if the constant  is fixed at
\begin{equation}
\textrm{Complexity} = \frac{1}{\pi \hbar}  \textrm{Action}.
\end{equation}
With this prefactor, an increase of complexity of a single gate corresponds to an advance of $e^{i\CA/\hbar}$ from $1$ to $-1$.
This translates to a rate of change of complexity of
\begin{equation}
\frac{d\CC}{dt} = \frac{2M}{\pi \hbar}.
\end{equation}
Neutral black holes precisely saturate the bound on the rate of change of complexity, Eq.~\ref{C-bound}. They saturate it whatever their size---small, intermediate or large compared to the AdS radius---and they saturate it whatever the number of spacetime dimensions.

In the original CV-duality, the answer for $d\CC/dt$ is not quite universal and would not allow such an interpretation in terms of the saturation of the  bound,  Eq.~\ref{C-bound}. For example, there are dimension-dependent factors that cannot be absorbed into a universal coefficient, and the  rate of change of action has some dependence on whether the black hole is small or large compared to the AdS length.

\subsection{Rotating BTZ black holes}

In \S\ref{sec:conserved-charge}, we saw that when there is a conserved angular momentum, the complexification bound tightens to
\be
\frac{d \CC }{dt} \leq \frac{2}{\pi \hbar} \left[ (M - \Omega J) -(M-\Omega J)_\textrm{gs} \right] .
\label{masterform}
\ee
Using Eq.~\ref{eq:BTZformulaforMJ}, the chemical potential for angular momentum is
\begin{equation}
\Omega = \frac{r_-}{r_+ \lads}.
\end{equation}
Rotating BTZ black holes thus have
\begin{align}
M-\Omega J=\frac{r_+^2+r_-^2}{8 G \lads^2}-\frac{2r_-^2}{8 G \lads^2}=\frac{r_+^2-r_-^2}{8 G \lads^2}=\sqrt{M^2-\frac{J^2}{ \lads^2}}.
\end{align}
With fixed $\Omega$, the ground state value of $M-\Omega J$ is $0$, given by $M = J =0$. Thus the ground state contribution  vanishes, and
\begin{equation}
(M-\Omega J)-(M-\Omega J)\Bigl|_{\text{gs}}=\sqrt{M^2-\frac{J^2}{\lads^2}}.
\end{equation}
In \S\ref{subsec-rotating} the rate of growth of action for the rotating BTZ black hole was computed
\begin{align}
\frac{d\CA}{dt}=\frac{r_+^2-r_-^2}{4G \lads^2}=2\sqrt{M^2-\frac{J^2}{\lads^2}}.
\end{align}
Thus rotating BTZ black holes precisely saturate the complexification bound in Eq.~\ref{CR-bound}.

\subsection{Small charged black holes} \label{sec:smallchargedblackholescomplexity}
We may also test our conjecture with electric charge. The bound of \S\ref{sec:conserved-charge} is now
\be
\frac{d \CC}{dt} \leq \frac{2}{\pi \hbar} \left[ (M - \mu Q) -(M-\mu Q)_\textrm{gs} \right] .
\label{masterform2}
\ee
We will now apply this formula to charged black holes that are much smaller than the AdS radius ($r_+ \ll \lads$), for which
\bea
f(r) &=& 1 -\frac{2GM}{r} + \frac{GQ^2}{r^2} = \frac{(r-r_+)(r-r_-)}{r^2}.
\eea
These have
\begin{equation}
GM = \frac{r_+ + r_-}{2}  \ \ \ \ \& \ \ \ \ GQ^2 = r_+ r_-.
\end{equation}
The chemical potential is
\be
\mu = \frac{Q}{r_+}.
\ee
For a given $\mu$, the smallest value of $M- \mu Q$ is zero (at leading semiclassical order), achieved by empty space $M=Q =0$. Thus the ground state contribution to \eqref{masterform2} vanishes and we find,
\bea
 (M - \mu Q) -(M-\mu Q)_\textrm{gs}  &=& \frac{r_+ + r_-}{2G} -\frac{Q^2}{r_+} = \frac{r_+ - r_-}{2G} = \sqrt{M^2 -\frac{Q^2}{G}} .
\eea
Small charged black holes exactly saturate the complexification bound in Eq.~\ref{C-bound}.

\subsection{Intermediate and large charged black holes}

\begin{figure}[h!]
\begin{center}
\includegraphics[width=4.5in]{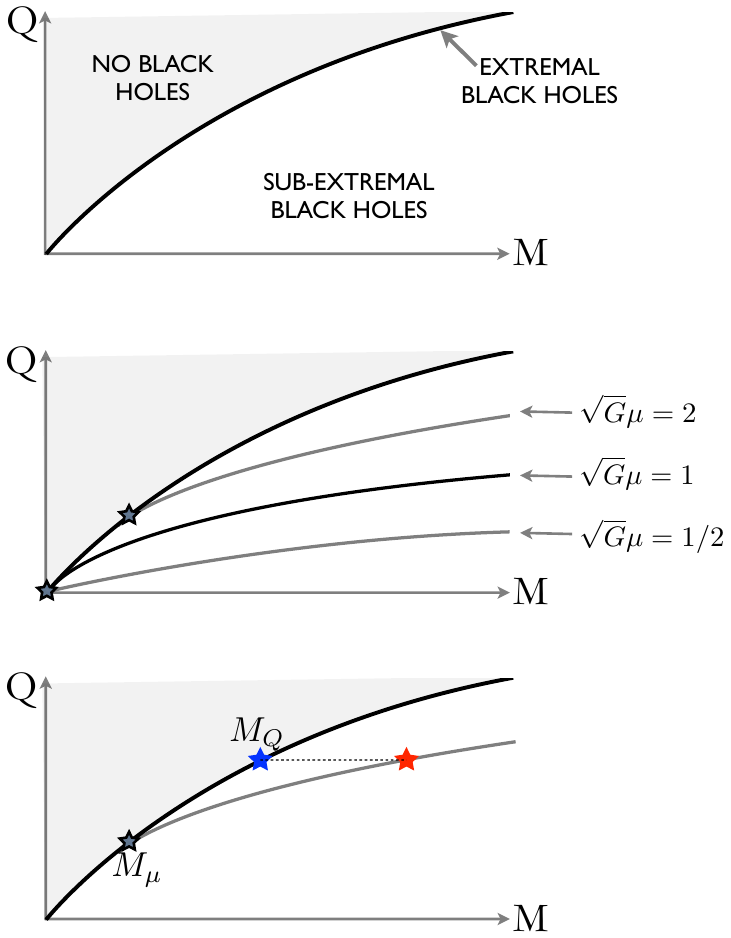}
\caption{The phase diagram for charged Reissner-Nordstrom black holes in AdS. {\bf Top pane}: at fixed $M$, black holes exist only for small enough $Q$. For  black holes that are small compared to $\lads$, the extremal line is $Q=\sqrt{G} M$; for black holes that are large compared to $\lads$, the extremal line becomes $Q \sim M^{1/3}$. {\bf Middle pane}: curves of constant chemical potential $\mu$. Small extremal black holes have $\sqrt{G} \mu = 1$; larger extremal black holes have larger $\mu$. Thus for $\sqrt{G}  \mu < 1$ the lines of constant $\mu$ end at $M = Q = 0$, and for $\sqrt{G}  \mu > 1$ the lines of constant $\mu$ end on the extremal line. {\bf Bottom pane}: for a given large charged black hole (red star),  we may define $M_\mu$ (gray star) as the mass of the extremal black hole with the same chemical potential $\mu$, and $M_Q$ (blue star) as the mass of the extremal black hole with the same charge $Q$.}
\label{fig-largechargedblackholes}
\label{extremal-curve}
\end{center}
\end{figure}

The situation for intermediate-sized  ($r_+ \sim \lads $) and large charged black holes ($r_+ \gg \lads $) is more complicated and leads to an apparent  violation of  the complexification bound Eq.~\ref{C-bound}. Let us consider the phase diagram in Fig.~\ref{fig-largechargedblackholes}. There are a number of important curves in the $(M,Q)$ diagram. The first  represents black holes at extremality. The extremality curve can be described by the parametric equations,
\bea
G^2M^2 \biggl|_\textrm{extremal} &=&\frac{\lads^2}{27}  (G\mu^2-1)(2G\mu^2 +1)^2 \cr \cr
Q^2 \biggl|_\textrm{extremal} &=& \frac{\lads^2}{3} \mu^2  (G\mu^2-1).
\eea
For small black holes it has the usual asymptotically flat form
\be
Q \biggl|_\textrm{extremal} =\sqrt{G}M.
\ee

The second class of curves shown in Fig.  \ref{fig-largechargedblackholes} are curves of constant chemical potential. The curves behave differently for $\sqrt{G}\mu<1$ and $\sqrt{G}\mu >1.$ The curve $ \sqrt{G}\mu=1 $  ends at $M=Q=0 $ where it is tangent to the extremal curve. For $\sqrt{G}\mu<1$ the constant-$\mu$ curves all end at $M=Q=0$, namely the AdS ground state. This is the minimum of $M - \mu Q$ for fixed $\mu < G^{-1/2}$.

For $\sqrt{G}\mu >1$ the constant-$\mu$ curves end on the extremal curve at some nonzero values of $M$ and $Q$. Where they intersect, the constant-$\mu$ curve and the extremal curve are tangent.  In these cases the nominal ground state at fixed  $\mu$ is the extremal black hole, which has $M - \mu Q$ negative.

\subsubsection*{The case $\sqrt{G}\mu<1$}
\begin{figure}[htbp] 
   \centering
   \includegraphics[width=3.3in]{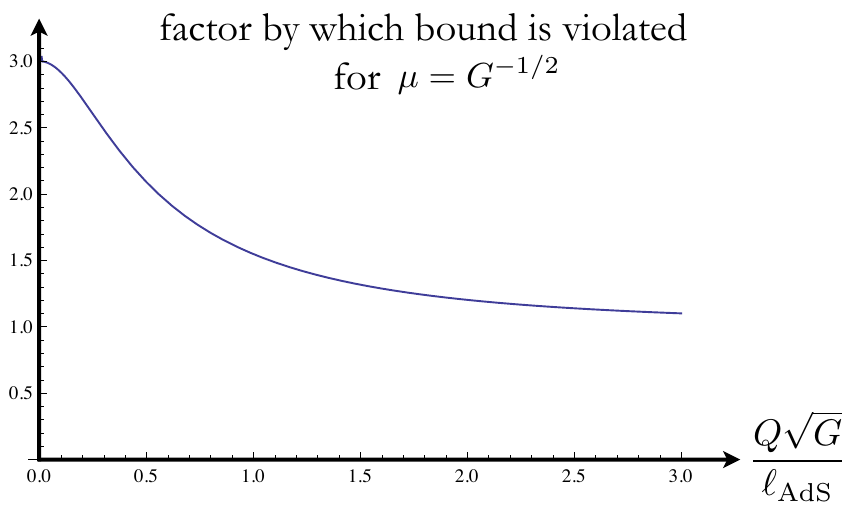}
   \caption{The rate of growth of action, Eq.~\ref{eq:chargedactionincrement}, divided by the bound of Eq.~\ref{eq:muis1bound}, along the line $\sqrt{G} \mu = 1$. The bound is violated everywhere along this line. For large $\mu =1$ black holes, the bound is violated by a small relative amount. For small $\mu = 1$ black holes, the black holes are almost extremal and  the bound is violated by a factor of 3. (Notice the subtle order of limits involved in keeping $\mu = 1$ for small black holes. If you first take $\lads \rightarrow 0$ then the bound is never violated, as in \S\ref{sec:smallchargedblackholescomplexity}: the extremality and ``small'' limits don't commute.)}
   \label{fig-violationformuis1}
\end{figure}

For $\sqrt{G}\mu<1$, the ground state is at $M=Q=0$ so that there is no apparent reason for a ground state subtraction. Comparing the action growth with the complexification bound, we find that for any mass and nonzero charge, the bound is violated. The violation gets worse for larger $\mu$, so the worst case in this regime is at $\sqrt{G}\mu=1.$  One finds that along this curve
\be
\sqrt{G} \mu = 1: \ \ \ \ \ 2(M-\mu Q)  - 2(M-\mu Q)_\textrm{gs} = 2(M-\mu Q) =  \frac{\sqrt{G}Q^3}{\lads^2} \label{eq:muis1bound}.
\ee
On the other hand the calculation of $\frac{1}{\pi \hbar}\frac{d\CA\textrm{ction}}{dt}$ in Eq.~\ref{eq:chargedactionincrement} exceeds the bound in \eqref{eq:muis1bound} by a modest factor. In Fig.~\ref{fig-violationformuis1} we show the ratio of the calculated rate of action growth to the bound given by \eqref{eq:muis1bound}. We see that for large $Q$ the ratio is close to unity, but that for small $Q$ there is a  significant $O(1)$ violation.  We will return to this apparent violation after considering the case $\sqrt{G} \mu>1.$

\subsubsection*{The case $\sqrt{G}\mu>1$}
In the case of large charged AdS black holes, there is a serious mismatch between the action calculation and the complexity expectation.  
The mismatch is most easily seen for very large black holes ($r_+ \gg \lads$) in the near extremal limit.

Consider very large ($r_+ \gg \lads$) charged black holes in $D=4$ spacetime dimensions, for which
\be
f(r)= \frac{r^2}{\lads^2} - \frac{2GM}{r} + \frac{GQ^2}{r^2} \ \ \ \& \ \ \  \mu = \frac{Q}{r_+}.
\ee
Extremality occurs at
\be
Q^2_\textrm{ext} = \frac{3\lads^2}{G}\left(  \frac{GM}{2\lads}      \right)^{4/3}=\frac{G\lads^2\mu^4}{3}.
\label{extr-curv}
\ee
The extremal $Q$ for a given $M$, Eq.~\ref{extr-curv},  defines a curve in the $M,Q$ plane  and is plotted in Fig.~\ref{fig-largechargedblackholes}.
%
%
%
%
%
Along the extremal curve the chemical potential is equal to the inverse slope. In other words, the contours of constant $\mu$ are tangent to the extremal curve,
\be
\frac{dM}{dQ} \biggl|_\textrm{extremal} = \mu.
\ee
After some straightforward calculations, one can see (assuming the ground state is given by the extremal black hole) that near extremality,
\be
(M-\mu Q)-(M-\mu Q)_\textrm{gs}=
2(M-M_Q) + O(M-M_Q)^2. \label{eq:mismatch11}
\ee

In Eq.~\ref{3.15}  \rm
we found that in that limit the rate of change of action is
\be
\frac{d \CA}{dt} = \sqrt{6} \sqrt{M_{Q}(M-M_{Q})} + O(M - M_Q) .
\label{eq:mismatch22}
\ee
Near extremality Eq.~\ref{eq:mismatch22} is much bigger than Eq.~\ref{eq:mismatch11}, apparently violating the bound. In this case the apparent relative violation is not by an $O(1)$ factor but becomes infinite as extremality is approached.

Thus the action proposal apparently violates the complexification bound. It is worth pointing out that this puzzle is not specific to the new CA proposal. It occurs in essentially the same form in the older CV proposal.

\subsection{Discussion}
\label{sec:A Pattern}

Of the four classes of black holes we have looked at, three worked perfectly. Neutral black holes (of any size and in any dimension), spinning BTZ black holes, and small charged black holes all had action growths that, when fed into our complexity equals action formula, gave rates of change of complexity that exactly saturate our bound.

Large charged black holes did not work. We found that a naive calculation of their rate of growth of action implied a complexification rate that was too large. While the rate satisfied the $2M$ bound of Eq.~\ref{C-bound}, it violated the tighter $2(M - \mu Q) - 2(M - \mu Q)_\textrm{gs}$ bound of Eq.~\ref{CQ-bound} that we conjectured should be satisfied by systems with conserved charge.

We believe we understand why the naive calculation of large charged black holes doesn't work, and why the other calculations do work. We suspect that it is to do with hair.

In theories with light charged particles, a ball of charge may form around a charged black hole. The hair becomes progressively more important as extremality is approached. At fixed $\mu$, this hair affects both the charged black hole and the ground state, and renders unreliable the calculation of the action, the calculation of $M - \mu Q$, and indeed the whole classical geometry. At fixed chemical potential $\mu > m/q$, the minimum of $M - \mu Q$ will feature a ball of charged particles (for $\mu > G^{-1/2}$ this ball will lie outside a black hole; for $G^{-1/2}>\mu >m/q$ there will be only a ball and no black hole).

Indeed, it should have been obvious that light charged particles will disrupt our analysis. Consider the claim that a black hole of mass $M$ and charge $Q$ is the fastest computer with that value of mass and charge.  In general this is obviously false.  Consider the case of an extremal black hole, which does not complexify at all; its complexification rate is zero. Let's compare that with a system consisting of a neutral black hole and a collection of electrons at a safe distance.  If we assume that the mass  and charge of the electron satisfy
\be
m< \frac{q}{\sqrt{G}}
\ee
then removing charge $Q$ in the form of electrons leaves the remaining neutral black hole with positive mass. This then does complexify, and therefore  charged black holes cannot generally be the fastest computers of a given mass and charge.

This explains why  our naive calculation of large charged black holes did not work. Now let's explain why small charged black holes  do work.  The reason is supersymmetry. In a sufficiently supersymmetric theory the BPS bound guarantees that there will be no light charged particles: every particle has $m \geq q /\sqrt{G}$. This means that sufficiently supersymmetric black holes are not susceptible to the formation of hair. Whenever our theory is a consistent truncation of a sufficiently supersymmetric theory, our calculations must be reliable.
Because flat space Reissner-Nordstrom emerges as the limit of the Strominger-Vafa \cite{StromingerVafa96}  $D1-D5$ system when the three charges are all equal \cite{CallanMaldacena96}, small Reissner-Nordstrom black holes had to work.

Similarly, rotating BTZ black holes had to work. A particle cannot have large angular momentum without having large energy; in addition, extremal BTZ black holes are known to have supersymmetric UV completions \cite{Coussaert:1993jp}.

On the other hand, large RN black holes in AdS did not have to work. We know of no examples in which they are sufficiently supersymmetric that our calculations had to be reliable. We had no reason to trust our naive calculations; and our naive calculations give us no reason to distrust our conjectures.

\subsection{Superconducting black holes}

In the last subsection we discussed that large charged black holes may be unstable to growing charged hair. In this subsection, we will look at an example in which this happens: holographic superconductors. We may ask how rapidly the thermal state of a superconducting black hole complexifies. The relevant hairy black holes correspond to far from extremal black holes of the type considered in \S\ref{sec:bh} with additional scalar condensate hair \cite{holosc1}. Our complexity growth calculation for far from extremal charged black holes leads to the action growth estimate $d\CA/dt \sim TS$, so we must compare $TS$ with  $(M - \mu Q) - (M - \mu Q)_{\text{gs}}$ to check the bound.

The zero temperature state of the holographic superconductor has all charge carried in the condensate and vanishing black hole radius and entropy \cite{holosc2}. Furthermore, the heat capacity of the superconducting black hole is a power law in $T$ at low $T$ \cite{holosc1}. The thermodynamic identity
\beq
(M - \mu Q) - (M - \mu Q)_{\text{gs}} = \int dT ( T \partial_T S)
\eeq
combined with $S(T) \sim T^\alpha$ implies that
\beq
(M - \mu Q) - (M - \mu Q)_{\text{gs}} = \frac{\alpha}{\alpha+1} TS.
\eeq
Since we estimated the action growth to be $d\CA/dt \sim TS$ it follows that the complexity growth bound is qualitatively obeyed. It would be interesting to make a more detailed calculation of the action so that a precise comparison with the complexity growth bound can be made.

In the next section we will discuss static shells surrounding black holes. This will provide another way to think about the superconducting condensate.

\section{Testing our conjecture with static shells}
\label{sec:staticshells}
\begin{figure}[h] 
   \centering
\hspace{2cm}    \includegraphics[width=3in]{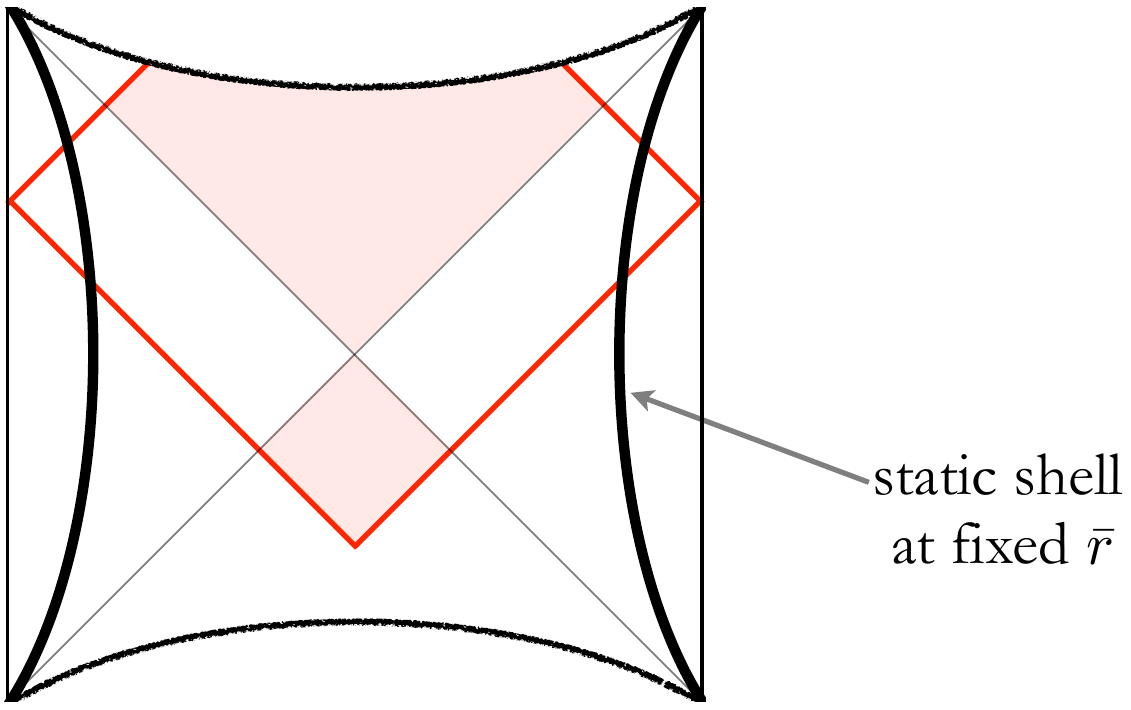}
      \caption{A black hole surrounded by a static  sphere.}
      \label{fig-sarcophagus}
\end{figure}

So far we have considered the complexification rate of an isolated black hole. We would like to test our proposal by adding controlled complications. One test is an ingoing shell (a shock wave), which will be considered in \S\ref{sec:shockwaves}; in this section we will instead consider a static shell (a Dyson sphere), see Fig.~\ref{fig-sarcophagus}.

Consider a static sphere at $r= \bar{r}$,  buttressed against collapse by its compressive strength. Outside the shell the metric is the Schwarzschild metric with mass $M + \delta M$,
\begin{equation}
ds^2  \Bigl|_{r> \bar{r}}=  - \left(1 - \frac{2G(M + \delta M)}{r}  + \frac{r^2}{\lads^2} \right) dt^2 + \frac{dr^2}{1 - \frac{2G(M + \delta M)}{r}  + \frac{r^2}{\lads^2} } + r^2 d \Omega_2^{\, 2 }.
\end{equation}
Inside the shell $r < \bar{r}$ the metric is the Schwarzschild metric with mass $M$ with a (locally unobservable) time dilation relative to infinity from being deep in the shell's gravitational well,
\begin{equation}
ds^2  \Bigl|_{r< \bar{r}} =  - {\frac{1 - \frac{2G(M + \delta M)}{\bar{r}}  + \frac{\bar{r}^2}{\lads^2}}{1 - \frac{2GM}{\bar{r}}  + \frac{\bar{r}^2}{\lads^2}}} \left(1 - \frac{2GM}{r}  + \frac{r^2}{\lads^2} \right) dt^2 + \frac{dr^2}{1 - \frac{2GM}{r}  + \frac{r^2}{\lads^2} } + r^2 d \Omega_2^{\, 2 }.
\end{equation}

Now let's calculate the action. The shell itself lies outside the horizon, and so only directly contributes to the time-independent (boost-invariant) part of the action. The only effect on the rate of change of action is indirect, through the change in the interior metric: the rate of change of action picks up exactly one gravitational time dilation factor,
\begin{equation}
\frac{d\CA}{dt_L} \Biggl|_\textrm{shell} = \sqrt{ {\frac{1 - \frac{2G(M + \delta M)}{\bar{r}}  + \frac{\bar{r}^2}{\lads^2}}{1 - \frac{2GM}{\bar{r}}  + \frac{\bar{r}^2}{\lads^2}}}} \frac{d \CA}{dt_L} \Biggl|_\textrm{no shell}.
\end{equation}
According to the action prescription, therefore, on the one hand, static shells do not compute, and, on the other hand, computers in gravitational wells slow down due to time dilation. Since both of these are exactly the behavior we would hope for, the proposal of Eq.~\ref{eq:CAformula} passes this test.

This argument applies to \emph{any} ordinary static matter outside the horizon---it does not compute to leading  order in $G$ and its only effect is that of gravitational time dilation, which is to slow the rate of computation of the black hole. This is consistent with the conjecture that an isolated  black hole is the fastest computer in nature.

Let us return now to the superconducting instability of charged black holes.
When the black hole becomes highly charged it will tend to expel charge into the atmosphere between the horizon and the AdS boundary. As discussed in Sec.\ref{sec:A Pattern}, the charge may then condense into a zero-entropy inert shell-like condensate. The remnant black hole at the center of the superconducting shell will not be close to extremality. Thus we will have an example of a far-from-extremal black hole surrounded by a static inert shell. The argument for static shells can be used to insure that the superconducting black holes do not violate the bound on computation.

\section{Testing our conjecture with shock waves}
\label{sec:shockwaves}

The original complexity/geometry duality passed a number of nontrivial tests involving the effects of shock waves on the growth of complexity and volume. It is important that the new CA-proposal passes the same tests.
In this section we consider the growth of complexity in eternal black hole geometries perturbed by shock waves and verify this.

Shock waves are constructed by perturbing the thermofield double state with thermal scale operators
\be
e^{-i H_L t_L}\, e^{-i H_R t_R}\, W(t_{n}) \dots W(t_1) \TFD, \label{eq:multiple-precursors}
\ee
where $W(t) = e^{i H_L t}\, W\, e^{-i H_L t}$, $W$ is a simple operator smeared over a thermal scale acting on the left boundary.\footnote{Here our conventions for precursors $W(t)$ are that Hamiltonian evolution $t_L$ increases to the future, but Killing time evolution in the left side of the bulk eternal black hole geometry increases to the past.} The state \eqref{eq:multiple-precursors} is dual to eternal black hole geometries perturbed by $n$ shock waves. Sometimes we will take $W$ to be smeared over the entire spatial boundary sphere. In these cases, we imagine that the field theory is sitting right above its Hawking-Page point so that smearing the operator over a thermal scale corresponds to smearing it over the sphere. At other times we will consider a high temperature limit with $W$ a spatially local operator smeared over the thermal scale. Similar states have been considered in the context of holography in \cite{shock,Shenker:2013yza,Leichenauer:2014nxa,localshocks,Shenker:2014cwa}, though they had previously been studied in other contexts \cite{Dray:1984ha,Dray:1985yt,Sfetsos:1994xa}.

These states provide a nontrivial check of the CA-duality. Because of boost invariance, in \S\ref{sec:bh} we only considered a one parameter family of states given by the overall time evolution $t_L + t_R$ of the thermofield double state. Shock wave states of the form \eqref{eq:multiple-precursors} let us dial the times $t_1 \cdots t_n$ and in some cases the spatial positions $x_1 \cdots x_n$ of an essentially arbitrary number of perturbations. Additionally, the tensor network construction of states \eqref{eq:multiple-precursors} gives specific predictions for the complexity growth \cite{localshocks} which is very naturally matched by the action prescription for complexity. These include matching growth rates that are different from linear in boundary time evolution $t_L$ \cite{localshocks} and matching cancellation effects between $e^{i H_L t}$ and $e^{-i H_L t}$ to the left and right of the $W$ operator \cite{complexityshocks}. While these tensor network predictions were all previously matched by the CV notion of complexity \cite{localshocks,complexityshocks,Susskind:2014jwa}, we will show they are just as naturally matched by the complexity equals action conjecture.

\subsection{One shock}\label{sec:one-shock}
First, we will consider the time evolved thermofield double state perturbed by a single precursor smeared over the entire boundary sphere
\be
e^{-i H_L t_L}\, e^{-i H_R t_R}\, W(t_w) \TFD. \label{eq:sinlge-precursors}
\ee
This state is dual to an eternal black hole geometry with a spherically symmetric shock wave emerging from the boundary at time $t_w$.
We will restrict to $t_w < 0$ and $t_L, t_R > 0$ so that we study the complexity of the state to the future of the shock wave.

The construction of these geometries has now been covered many places (see, e.g. \cite{localshocks}), so we will give only a minimal review. We will consider perturbations of global-AdS black hole geometries in $D$  dimensions. It is convenient to consider Kruskal coordinates, where the metric is given by
\begin{align}
ds^2 =\,\,-A&(uv) \, 2du dv + B(uv)\, d\Omega_{D-2}^2 + 2A(uv)h\,\delta(u) \, du^2, \\
&A(uv) = -\frac{2}{uv}\frac{f(r)}{f'(r_h)^2}, \qquad B(uv) = r^2, \\
&f(r) = 1 - \frac{8 \pi}{(D-2)\Omega_{D-2}} \frac{2 G M}{r^{D-3}} + \frac{r^2}{\lads^2}, \label{eq:metric-one-shock}
\end{align}
where $r_h$ is the horizon radius, and the last term is an ansatz for the backreaction of the perturbation.
The relationship between Kruskal coordinates and Schwarzschild coordinates is given by
\be
uv = - e^{\frac{4\pi}{\beta}r_*(r)}, \qquad u/v = - e^{-\frac{4\pi}{\beta} t}, \qquad \beta = \frac{4\pi \lads^2}{r_h (D-1) + (D-3)\frac{\lads^2}{r_h} },
\ee
with $dr_* = dr / f(r)$.

The shock wave is created by acting with the scalar operator $W$ at time $t_w<0$ smeared over the entire boundary. For large $|t_w|$, this creates a particle of null matter traveling along $u=0$ in the bulk. The expectation value of the stress tensor in this state is
\be
T_{uu} = \frac{E}{\lads^{D}} e^{ 2\pi |t_w| / \beta } \delta(u), \label{eq:stress-tensor-shock}
\ee
with the dimensionless constant $E$ related to the $O(1)$ asymptotic energy of the particle (see e.g. the Appendix of \cite{Roberts:2014ifa} for details). Plugging \eqref{eq:metric-one-shock} and \eqref{eq:stress-tensor-shock} into Einstein's equations,
we find a solution
\be
h \sim  e^{\frac{2\pi}{\beta}(|t_w| - t_* ) } \label{eq:shift}
\ee
where we have defined the fast scrambling time $t_* = \frac{\beta}{2\pi} \log \frac{\lads^{D-2} } {G}$. The main point is that since boundary time evolution acts as a boost near the horizon, the $G$ suppression can be overcome by pushing $t_w$ further into the past. In this way, a small asymptotic perturbation can have a large backreaction on the geometry and a large effect on the complexity of the state.
Finally, we note that the metric \eqref{eq:metric-one-shock} implies that the constant $h$ has the interpretation of a shift in the $v$ coordinate by $\delta v = h$ from crossing the shock. This means, among other things, that the left and right $v=0$ horizons no longer meet at $u=0$.

\begin{figure}
\begin{center}
\includegraphics[scale=.61]{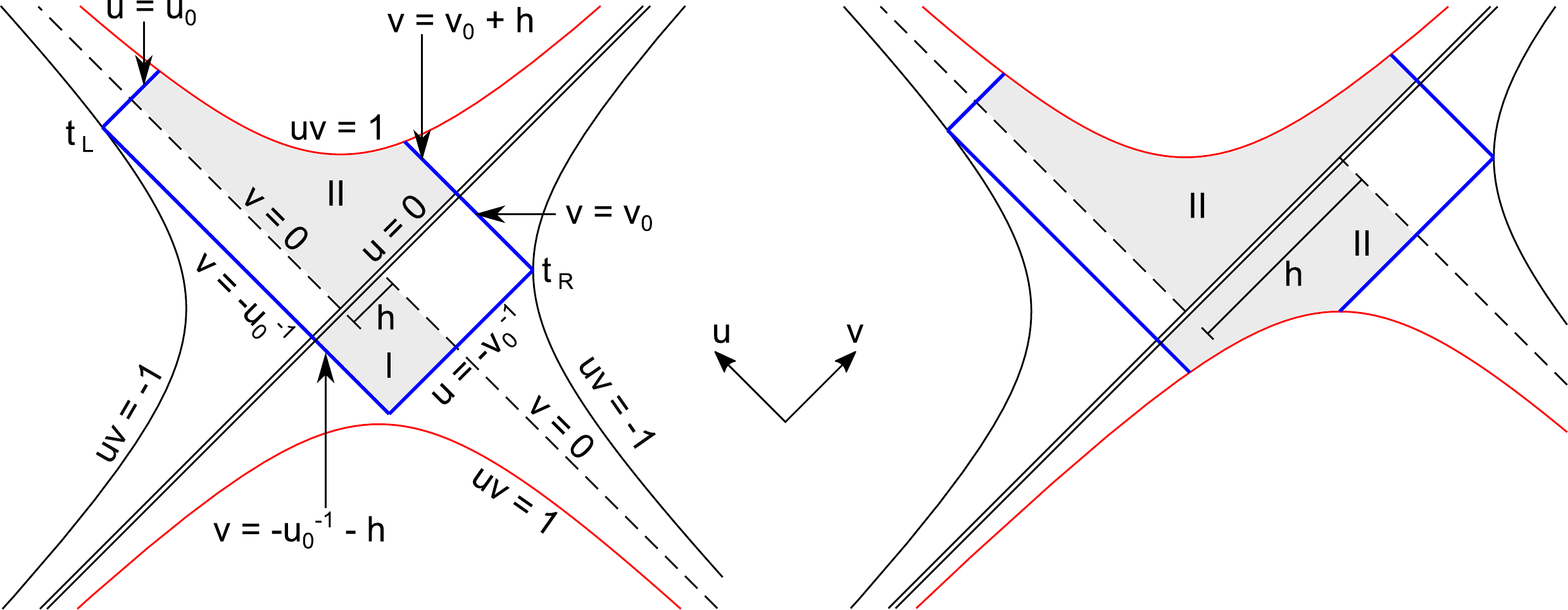}
\caption{Kruskal diagram of one-shock geometries with different size shifts. The double black line along $u=0$ is the shock wave. The blue lines show the boundary of $\CW$. The light gray shaded region is the intersection of $\CW$ and the black hole interior. The action in this gray region determines the complexity.
{\bf Left:} small shift $h$ with $|t_w| - t_* \le t_R$. {\bf Right:} large shift $h$ with $|t_w| - t_* \ge t_R$ .}\label{kruskal}
\end{center}
\end{figure}

The Kruskal diagram of this geometry is shown in Fig.~\ref{kruskal}.\footnote{Fig.~\ref{kruskal} is only really appropriate for $D=2+1$. For $D>3$, the left and right boundaries of the Kruskal diagram are no longer at $uv=-1$ \cite{Fidkowski:2003nf}.  Since this complication does not change the asymptotic value of the complexity of the state, it will be neglected.} The left figure shows a geometry with a small shift $h$, and the right figure shows a larger shift. The shock wave is shown as double black lines along $u=0$.  The Wheeler-DeWitt patch $\CW$ is drawn in blue, and the intersection of $\CW$ and the black hole interior shown is light gray.\footnote{In this section, we will be interested only in the intersection of $\CW$ and the Einstein-Rosen bridge. The portion of $\CW$ outside of the black hole interior is time independent and divergent. Therefore, dropping this region is a consistent regularization of the complexity for all of the states that we consider.}  The shape of this region (and whether it touches the bottom singularity) is determined by the boundary times $t_L, t_R$ and the shift $h$. Geometrically, we see that for $u_0^{-1} + h  < v_0$ the past going light rays meet in the interior, and for $u_0^{-1} + h  > v_0$ they intersect the singularity. Here, we have used the definition of the Kruskal coordinates to define
\be
u_0 = e^{\frac{2\pi}{\beta} t_L}, \qquad v_0 = e^{\frac{2\pi}{\beta} t_R}. \label{eq:boundary-times-as-kruskal}
\ee
Since we are assuming $t_L > 0$, $u_0^{-1}$ is exponentially small and from now on will be neglected. This lets us express the condition for $\CW$ to intersect the past singularity as $|t_w| - t_* \ge t_R$.

\begin{figure}
\begin{center}
\includegraphics[scale=.7]{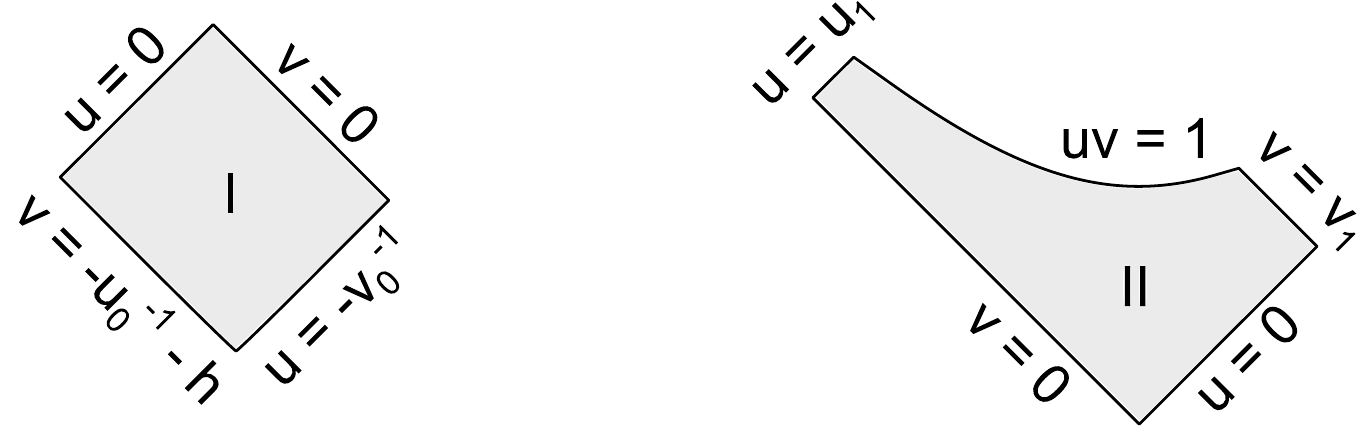}
\caption{{\bf Left:} diamond-shape region bounded on the top by the horizons and on the bottom by edges of $\CW$. This region will always have vanishing action. {\bf Right:} five-sided region bounded on the bottom by the horizons, on the sides by edges of $\CW$, and on the top by the singularity. The only significant contribution to action is from regions with this shape.}\label{regions}
\end{center}
\end{figure}

The complexity will be a function of $u_0, v_0,$ and $h$, since these are the parameters of the state \eqref{eq:sinlge-precursors} in reference to the thermofield double. The geometry of the causal wedge is different depending on the value of these parameters, with a transition at $|t_w| - t_* = t_R$ or $h  = v_0$.  We expect that the complexity $\CC$ will be a piecewise continuous function due to the transition of the geometry of the $\CW$ at the point $h  = v_0$. Using the results from \S\ref{sec:bh}, we could easily compute the rate of change of complexity for each of $t_L, t_R, t_w$ and then integrate to get the overall complexity. Instead, we will take a slightly different approach. From Fig.~\ref{kruskal}, we see that either the intersection of $\CW$ with the black hole interior creates a diamond-shaped region bounded on two sides by the horizons and on two sides by the Wheeler-DeWitt patch or it creates a five-sided region bounded by the two horizons, the two edges of the patch, and the singularity. These two shapes, denoted $I$ and $II$ in Fig.~\ref{kruskal}, are shown in Fig.~\ref{regions}. Since $v_0^{-1} =  e^{-2 \pi t_R /\beta} \ll 1$, the diamond region $I$ is exponentially small \eqref{eq:boundary-times-as-kruskal} and will have a vanishing contribution to the action. The only important contribution to the action will come from the five-sided region $II$ bounded by the singularity.

 To proceed, we will compute the action of region $II$ for arbitrary $\CW$ edges $u_1, v_1$. Since $t_L, t_R, |t_w| > 0$, we can safely assume $u_1, v_1 \gg 1$. The bulk contribution is straightforward, and we find
 \be
\CA_\textrm{bulk} = -\frac{\Omega_{D-2} \, r_h^{D-1}}{ \pi G \lads^2} \frac{\beta}{2\pi} \log u_1 v_1.
 \ee
 The boundary contributions are from the horizon and the singularity. The edges of $\CW$ do not contribute. A calculation similar to that in \S\ref{sec:bh} gives
  \be
\CA_\textrm{boundary} =  \left[  - \bigg( \frac{D-1 }{D-2}\bigg) M  + \frac{\Omega_{D-2} r^{D-3} }{8 \pi G}  \left( (D - 2 ) + (D-1) \frac{r^2}{\lads^2} \right) \right]^{r_h}_0  \frac{\beta}{2\pi} \log u_1 v_1,
 \ee
 and the total action in region $II$ is
 \be
\CA_{II} = 2M\, \frac{\beta}{2\pi} \log u_1 v_1. \label{eq:action-region-II}
 \ee

With this result in hand, let us consider the one shock geometry with small shift $|t_w| - t_* \le t_R$ (left side of Fig.~\ref{kruskal}). The only contribution is from the top five-sided region $II$. We can simply plug $u_1 = u_0$ and $v_1 = v_0 + h$ into \eqref{eq:action-region-II}. Using the fact that $v_0 > h$, we find
\be
\CA_{|t_w| - t_* \le t_R} = 2M\, (t_L + t_R),\label{eq:one-shock-complexity-weak-shock}
\ee
that is, as if the shock wave wasn't even there. This is easy to understand; for large $t_R$ we have evolved the state to a region where the backreaction of the perturbation is negligible. The complexity is simply given by the time evolution of the thermofield double state.

Now, let's consider the geometry with large shift $|t_w| - t_* \ge t_R$ (right side of Fig.~\ref{kruskal}). Now we have two five-sided region $II$s. We get the contribution from the top region as before by  plugging $u_1 = u_0$ and $v_1 = v_0 + h$ into \eqref{eq:action-region-II}. Similarly, we get the contribution from the bottom region by again using \eqref{eq:action-region-II}, but with  $u_1 = -v_0^{-1}$  and $v_1 = -u_0^{-1} - h$. (It's OK to flip it upside down.) This time using the fact that $v_0 > h$ and $u_0^{-1} \approx 0$, we find
\be
\CA_{|t_w| - t_* \ge t_R} = 2M\, \big[t_L - t_R + 2(|t_w| - t_*)\big].\label{eq:one-shock-complexity-strong-shock}
\ee
In this case, the shift $h$ is large, the backreaction of $W(t_w)$ is not negligible, and the perturbation makes a huge contribution to the complexity. Furthermore, the growth of complexity in $t_w$ is twice the rate of growth in $t_L$. We can think of this as coming from the fact that $W(t_w) = e^{i H_L t_w} W e^{-i H_L t_w}$ is made up of two time evolution operators each of which accrues complexity linearly with time.

Additionally, we see that the growth of complexity in $t_w$ is delayed by a scrambling time $t_*$. From the bulk perspective, this is the fact that we need to push the time of the shock very far into the past in order to overcome the $G$ gravitational suppression. From the boundary/complexity perspective, this has been called the  ``switchback'' effect \cite{complexityshocks}. If $W$ were the identity operator, the two time evolution operators on either side of $W(t_w)$ would cancel. $W$ is not the identity; it is a very simple operator (e.g. a perturbation of only one of the $N^2$ degrees of freedom in a lattice site). While the perturbation is still growing, there will be cancellation between the two time evolution operators. Fast scrambling dynamics dictate that the influence of this perturbation will grow exponentially, covering all $N^2$ degrees of freedom in a time $t_* = \frac{\beta}{2\pi}\log N^2$. Since the growth is exponential, almost none of the degrees of freedom are influenced until the very end. Thus, there is a delay of $t_*$ until the complexity can grow. This was predicted and matched \cite{complexityshocks} for  CV-duality, and here we are showing how simply it arises for the complexity equals action prescription.

In the next subsections, we will give extensions and related results for complexity growth in shock wave geometries. However, many of the details of these calculations will be left to the references and/or the reader.

\subsection{Finite energy shocks}
In the limit studied in the last section, the time of the shock $|t_w|$ was taken to infinity, and the asymptotic energy of the shock (as compared to the mass of the black hole) $E/M$ was taken to zero, such that the quantity $h \sim e^{2\pi t_w / \beta} E/M$ was held fixed. This was to ensure that the precursor was a small perturbation and provided a simple relationship between the coordinates when crossing the shock. A relic of this limit was that the shock wave had to lie on the $u=0$ horizon. Additionally, the rate of complexity growth was always fixed at $2M$. If instead we inject finite energy into the system from the left boundary, we expect that the complexification rate should change accordingly.

\begin{figure}
\begin{center}
\includegraphics[scale=.8]{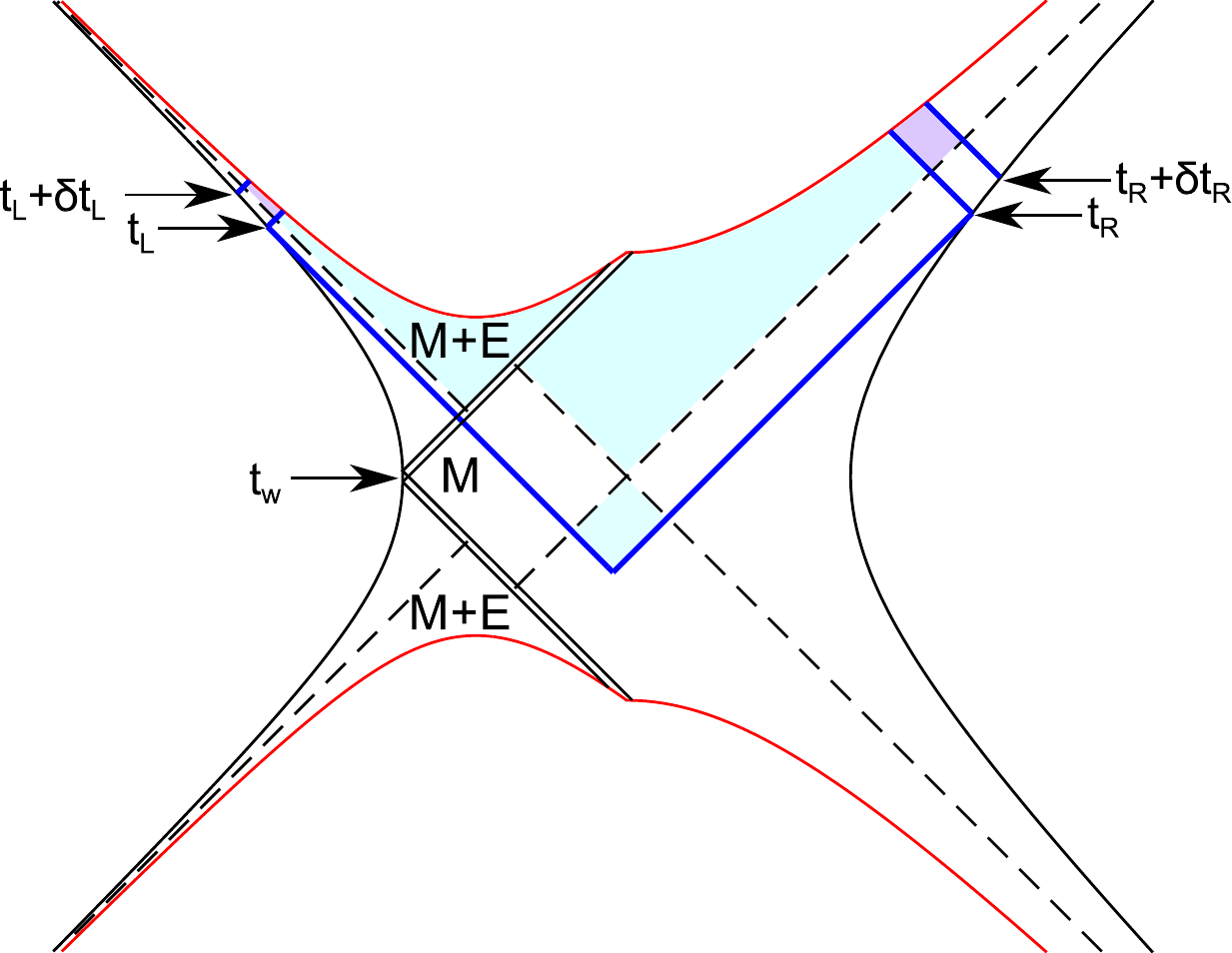}
\caption{Complexity growth with finite time perturbation at $t_w = 0$ with finite energy $E$. The change in complexity is given by the purple shaded regions. The $t_L$ evolution will grow at rate $2(M+E)$, but $t_R$ evolution will grow at rate $2M$.}\label{finite-complexity}
\end{center}
\end{figure}

Let us study precursors of finite energy $E/M$ and at finite $u$. An example of such a geometry is shown in Fig.~\ref{finite-complexity}.  The details of the construction of this geometry are left to Appendix~\ref{appendix-finite-time}. The picture for the state \eqref{eq:sinlge-precursors} is that of the perturbation emerging from the past singularity, materializing on the boundary at time $t_L=t_w$, and then traveling off into the future singularity. The past horizon shrinks by some amount after ejecting the perturbation, and the future horizon grows outward by a different amount after swallowing the perturbation. These are the analogs of the simple shifts usually considered in the infinite time shock wave geometries, see Appendix~\ref{appendix-finite-time}. Importantly, we think of the state \eqref{eq:sinlge-precursors} in the Schr\"{o}dinger picture. The perturbation is always present, and $t_w$ is simply a label that tells us when the perturbation materializes on the boundary.

From the figure, it's clear that the increase in $t_R$ is solely in the region of spacetime with energy $M$. On the other hand, time evolution on the left boundary results in an addition to the Wheeler-DeWitt patch that is solely in the region of spacetime with energy $M+E$. Thus, we find
\be
\frac{d\CA}{dt_R} = 2M, \qquad \frac{d\CA}{dt_L} = 2(M+E), \label{eq:complexity-finite-time-shock}
\ee
which is appropriate since we injected energy $E$ into the left CFT. Even if we had looked at really early times $t_L, t_R \ll -\beta$ such that $\CW$ starts from the bottom of the Kruskal diagram, we would have still found the same rate of growth. This is indicative of the fact that the perturbation is always present in the left CFT. The energy is always $M+E$, and so the complexification rate should reflect that.

\subsection{Multiple shocks}

\begin{figure}
\begin{center}
\includegraphics[scale=.85]{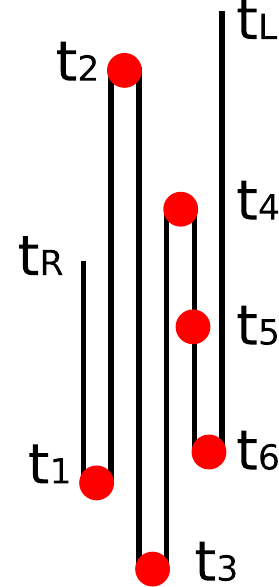}
\caption{A time fold with six operator insertions. This time fold depicts the construction of the out-of-time order state with six precursor insertions: $e^{-i H_L t_L}\,  W(t_6)W(t_5)W(t_4)W(t_3)W(t_2)W(t_1)\, e^{-i H_R t_R}\TFD$, where all the operators are understood to act on the left.
In this figure, all of the insertions except for $t_5$ occur at switchbacks.}\label{fig-time-fold}
\end{center}
\end{figure}

Next, let us comment on states perturbed by multiple precursors as in \eqref{eq:multiple-precursors}. As before, we will consider a compact boundary theory but this time perturbed $n$ spherically symmetric shock waves described by shifts $h_i\sim e^{ -\frac{2\pi}{\beta}(t_* \pm t_i )}$, with a ``$+$'' for $i$ odd and a ``$-$'' for $i$ even. We will also assume that all the shocks are strong, $h_i \gg 1$.  One can easily construct these geometries following the recipe of \cite{Shenker:2013yza}. In \cite{complexityshocks}, it was found that the total complexity of the state, using  CV-duality, is given by
\be
\CC \sim t_f - 2n_{sb}t_*, \qquad \text{(CV-duality)},
\ee
where $t_f$ is the length of the total time fold of the state, and $n_{sb}$ is the number of operators that are inserted at switchbacks.
A time fold is a convenient depiction of the construction of an out-of-time order state, such as \eqref{eq:multiple-precursors}. The total length of the time fold $t_f$ was defined in \cite{complexityshocks} as
\be
t_f \equiv |t_1-t_R| + |t_2 - t_1| + \dots |t_L - t_n|,
\ee
and it was assumed that $|t_i - t_{i+1}| > t_*$. We already discussed switchbacks in \S\ref{sec:one-shock}. From the perspective of the time fold, a switchback is an operator insertion where the time fold ``folds'' back on itself. An example time fold is shown in Fig.~\ref{fig-time-fold}. An insertion could fail at a switchback if the $(i+1)$th precursor in the state \eqref{eq:multiple-precursors} cancels against the time evolution of the $i$th precursor. See \cite{complexityshocks} for more details.

\begin{figure}
\begin{center}
\includegraphics[scale=.70]{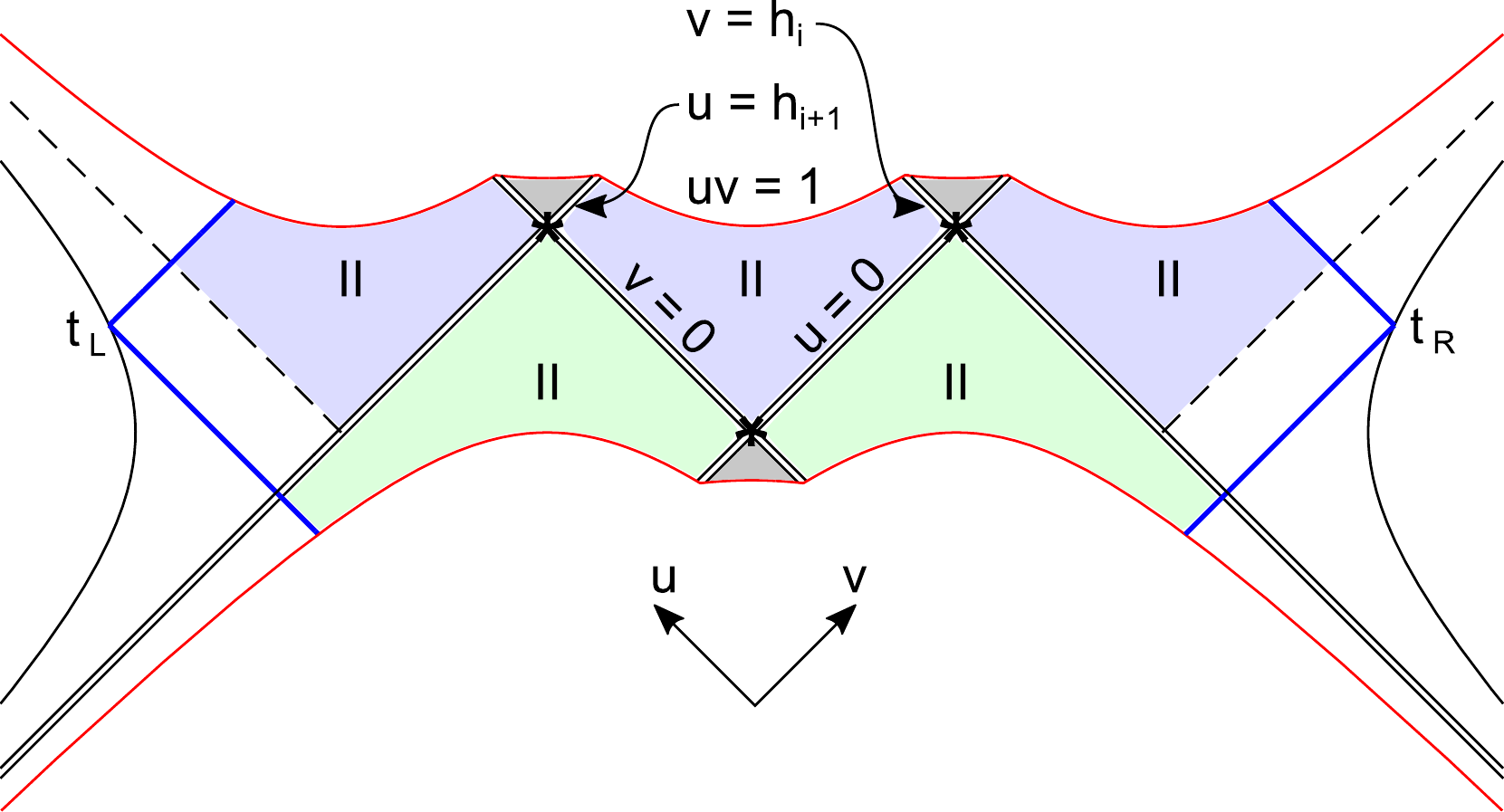}
\caption{Kruskal diagram of a geometry perturbed by four shock waves. The blue lines show the boundary of $\CW$, the shocks are drawn with double black lines, and the stars show their collisions. The intersection of $\CW$ and the Einstein-Rosen bridge is colored blue, green, and gray. The dominant contribution to the action is given by the five blue and green regions labeled $II$. The gray ``postcollision'' regions have negligible action. }\label{fig-multiple-shocks}
\end{center}
\end{figure}

It's easy to see, following the procedure outlined \S\ref{sec:one-shock}, that the CA-duality reproduces this result. Let us prune the list of precursors so that only those at switchbacks are represented in the state \eqref{eq:multiple-precursors}. An example geometry with four shocks is shown in Fig.~\ref{fig-multiple-shocks}.
The gray ``postcollision'' regions are complicated and cannot be described by the $(u,v)$ coordinates. (For more details, see \cite{Shenker:2013yza}.) However, if the strengths of all of the shocks are strong $h_i \gg 1$, these regions become small and have vanishing action. The only contribution will be from the five-sided region $II$s. With $n$ shock waves, there will be $n+1$ five-sided region $II$s contributing to the action. The contribution of each one is given by \eqref{eq:action-region-II}, and the $i$th one (counting from the right) gives
\be
\CA_{i} = 2M\, \frac{\beta}{2\pi} \log h_{i-1} h_{i},
\ee
where we have abused notation so that $h_0 = v_0$, $h_{n+1} = u_0$, we have assumed $t_L, t_R >0$, and we have assumed an even number of shocks for simplicity. Summing over all regions, we get precisely that the action of the Wheeler-DeWitt patch grows as
\be
\CA_{n\textrm{\,shocks} } = 2M (t_f - 2n_{sb}t_*) \qquad \text{(CA-duality).}
\ee

\subsection{Localized shocks}
\begin{figure}
\begin{center}
\includegraphics[scale=.7]{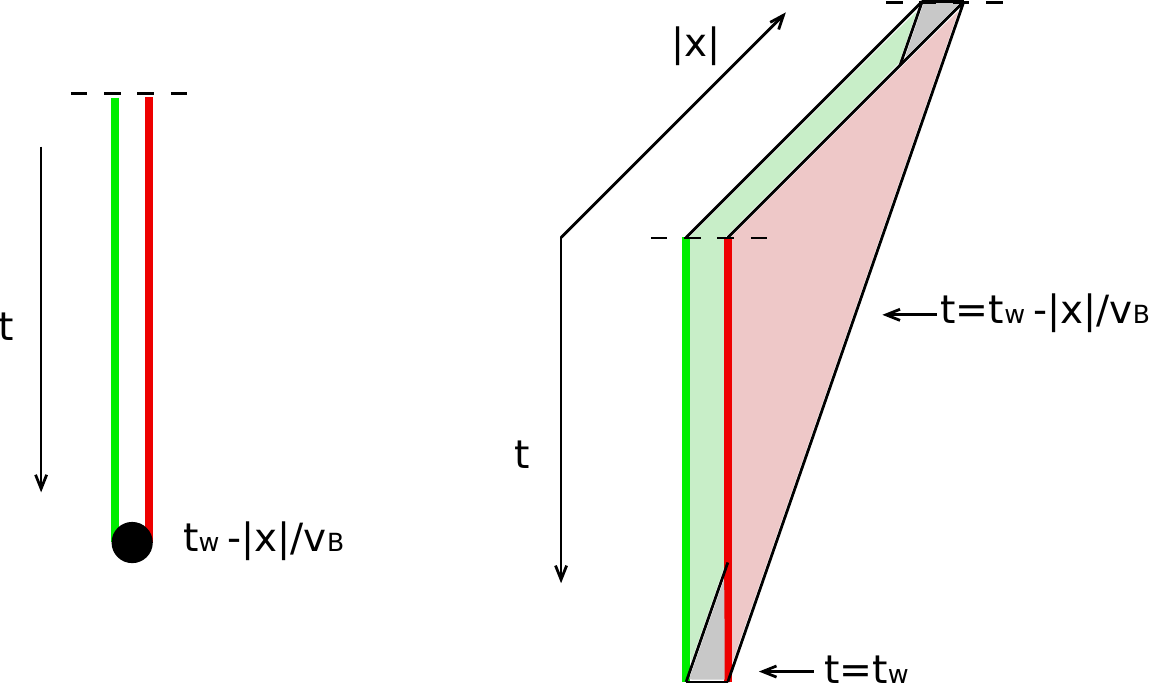}
\caption{A time fold for the state \eqref{eq:sinlge-local-precursors} with $t_L=t_R=0$. The length of the fold is dependent on position. The minimal tensor network for the state is given by fibering the time fold over the transverse space.}\label{fig-localized-time-fold}
\end{center}
\end{figure}

Another generalization is to localize the precursor operators in the transverse space. In the context of scrambling these geometries were introduced in \cite{shock} and studied extensively in \cite{localshocks}. The $D=2+1$ case for a single localized precursor was also studied in purely two-dimensional CFT terms in \cite{Roberts:2014ifa}. Such a state is given by
\be
e^{-i H_L t_L}\, e^{-i H_R t_R}\, W_x(t_w) \TFD, \label{eq:sinlge-local-precursors}
\ee
where as usual $W_x(t_w) = e^{i H_L t_w}\, W_x\, e^{-i H_L t_w}$, and $W_x$ is a simple thermal scale operator localized on the boundary at spatial position $x$. In this case, the time fold of the operator insertion is position dependent. This is shown in Fig.~\ref{fig-localized-time-fold} for the state \eqref{eq:sinlge-local-precursors} with $t_L=t_R=0$. This also means that the strength of the shock wave dual to the operator insertion has a position dependent strength $h(x) \sim e^{\frac{2\pi}{\beta}(|t_w| - t_* - \mu|x|) }$, where $\mu$ is a positive constant and $|x|$ is the transverse distance from the shock. For more information, see \cite{localshocks}.

\begin{figure}
\begin{center}
\includegraphics[scale=.8]{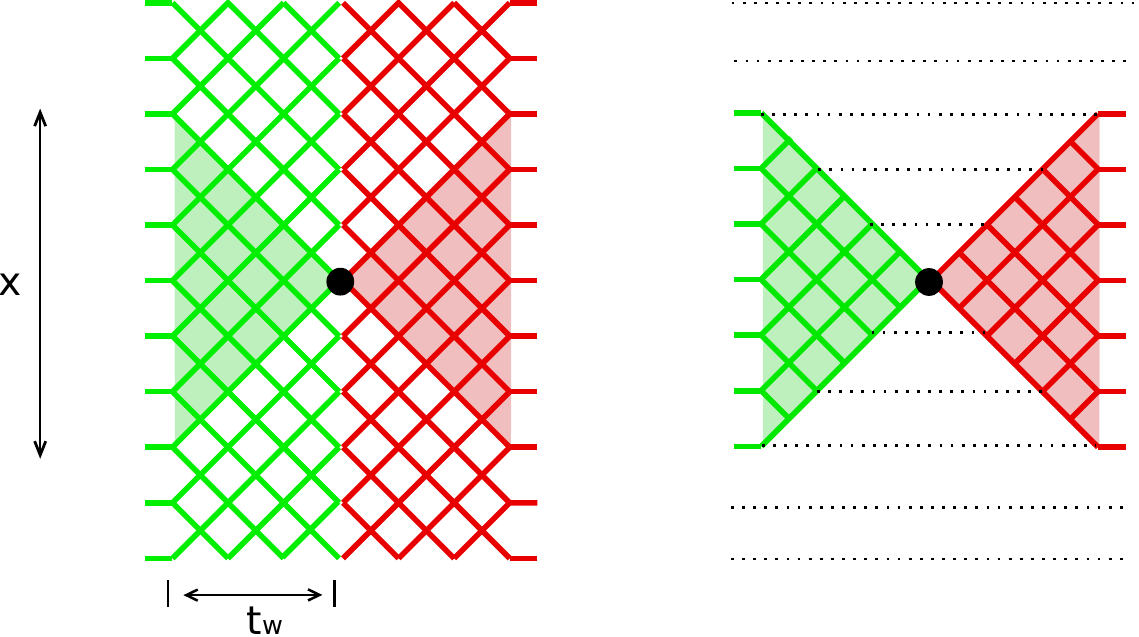}
\caption{A tensor network for the operator $W_x(t_w) = e^{i H_L t_w}\, W_x\, e^{-i H_L t_w}$. The red represents the evolution $e^{-i H_L t_w}$, the black dot represents the local insertion $W_x$, and the green represents the reverse evolution $e^{i H_L t_w}$. {\bf Left:} naive tensor network before cancellation. {\bf Right:} tensor network after cancellations outside the influence of the insertion $W_x$.}\label{fig-localized-tensor-network}
\end{center}
\end{figure}

There are a few interesting differences in the complexity growth of the localized one-shock state and the spherically symmetric one-shock state. The early $|t_w|$ complexity growth is no longer linear in $t_w$.  This is because the operator can grow in spatial directions as well as the $N^2$ ``matrix'' directions, so the switchback effect now includes cancellations in the transverse space. A tensor network for a single localized precursor is shown in Fig.~\ref{fig-localized-tensor-network}. The precursor will grow ballistically with characteristic velocity $v_B$, which is known as the butterfly velocity \cite{shock,localshocks}. This velocity can be computed holographically for Einstein gravity, and for high temperatures is given by the dimension dependent formula \cite{shock,localshocks}
\be
v_B = \sqrt{\frac{D-1}{2(D-2)} }.
\ee
This spatial growth means that the tensor network Fig.~\ref{fig-localized-tensor-network} has the geometry of two solid cones. By considering the point (as a function of transverse coordinate $x$) where the boundary of $\CW$ moves off the singularity (e.g. the point $h(x)=v_0$ in Fig.~\ref{kruskal}) as a way of defining the geometry of the Wheeler-DeWitt patch, we find an exact match between the geometry of the tensor network shown in Fig.~\ref{fig-localized-tensor-network} and the geometry of $\CW$ inside the Einstein-Rosen bridge.

On the other hand, if the spatial topology of the field theory is still compact, the precursor will eventually run out of room to grow. Once this saturation occurs, the growth with $|t_w|$ will transition to being linear. However, in this case there is an additional delay in the growth as compared to \eqref{eq:one-shock-complexity-strong-shock} related to the size of the spatial boundary and time it takes for the operator to saturate. (Se, e.g. the caption in Fig.~11 in \cite{localshocks}.)

The spatially compact boundary situation is slightly complicated by the fact that the shock will collide with itself on the other side of the transverse sphere. Instead, let us consider the planar-AdS black hole geometry. Since the strength of the shock is position dependent, the action in $\CW$ is no longer simply linear in $t_w$. Let us match this action to the volume of the tensor network describing the state. Let $m$ be the energy density, $L$ the infinite length of a transverse direction, and $M = mL^{D-2}$ the total energy. Using the techniques of \S\ref{sec:one-shock}, it's not hard to show that the total action is
\be
\CA = 2M(t_L + t_R) +  \frac{4 m v_B^{D-2} (|t_w| - t_* - t_R)^{D-1}}{(D-1)(D-2)    }, \label{eq:one-precursor-complexity}
\ee
which is consistent with both the result from  CV-duality and the tensor network predictions of \cite{localshocks}. In particular, with $t_L = t_R=0$, the second term is exactly the volume of the two solid $D-1$-dimensional cones shown in Fig.~\ref{fig-localized-tensor-network}.

One can also consider states perturbed by multiple localized precursors
\be
e^{-i H_L t_L}\, e^{-i H_R t_R}\, W_{x_n}(t_{n}) \dots W_{x_1}(t_1) \TFD. \label{eq:multiple-localized-precursors}
\ee
The analysis is annoyingly complicated and does not provide any additional insight into the complexity equals action conjecture. Instead, we will simply state that all of the results from  CV-duality from \cite{localshocks} carry over. We can construct the tensor network by fibering the time fold for the state over the transverse directions \cite{localshocks}. As mentioned, we can determine the geometry of $\CW$ by considering the point (as a function of transverse coordinate $x$) where the boundary of $\CW$ moves off the singularity. This lets us find an exact match between the geometry of the tensor network that constructs the state \eqref{eq:multiple-localized-precursors} and the geometry of $\CW$ inside the Einstein-Rosen bridge.

\subsection{Comment on transparency}

Finally, let us comment on the transition in complexity growth that occurs at $|t_w| - t_* = t_R$. This transition is precisely the transparent and opaque horizons of \cite{opacity}. From \eqref{eq:one-shock-complexity-strong-shock} we see that there is a period during which the complexity decreases on the right side,
\be
\frac{d\CC}{dt_R} < 0, \qquad (|t_w| - t_* > t_R).
\ee
This corresponds to the thermodynamically rare situation of an ``opaque'' horizon. Any observer that jumps into the black hole will get destroyed at the horizon by a high energy shock wave!

On the other hand, for $t_R > |t_w| - t_*$
 \eqref{eq:one-shock-complexity-weak-shock} indicates that the complexity resumes it normal increasing behavior,
\be
\frac{d\CC}{dt_R} > 0, \qquad (|t_w| - t_* < t_R).
\ee
According to \cite{opacity} increasing complexity
corresponds to a ``transparent'' horizon. As mentioned in \S\ref{sec:one-shock}, this is because we have evolved the state to a point where the backreaction of the precursor is negligible. Were an observer to jump in, at the horizon he or she would only encounter a harmless particle of vanishing energy.

We can also use the CA-duality to generalize the notion of transparency to localized shock wave geometries. For these geometries, the strength of the shock wave $h(x)$ is dependent on spatial position.
One can define a specific complexity $c(x)$ by not integrating the action over the spatially transverse directions of the Wheeler-DeWitt patch. This is roughly a measure of how the complexity is changing along these transverse  directions. We expect that $c(x)$  will be a piecewise continuous function, with the point of transition depending on $x$. As before, the transition occurs when $\CW$ moves off the past singularity. Computing this point of transition as a function of $x, t_w, t_R$, we find that a ``wave of transparency'' propagates ballistically with the butterfly velocity $v_B$.

\section{Tensor network model of\\ Einstein-Rosen bridge growth}
\label{sec:tensornetworks}

In this section we will perform a nontrivial test of our conjecture using tensor networks. Tensor networks provide a microscopic account of complexity growth, and we will show that this microscopic account  agrees with the results from CA-duality. Further, as the case with shock waves in \S\ref{sec:shockwaves}, this derivation highlights that the complexity growth is dual to the \emph{minimal} quantum circuit that builds the state.

The tensor-network model serves several purposes. It justifies the reduction of the UV Hilbert space of the CFT to a number of active qubits that is not UV sensitive and instead scales like the entropy; it gives a concrete model of complexity along the lines discussed in Appendix~\ref{sec:boundapp}; it shows that the growth of complexity is roughly given by $TS$; it justifies the Hartman-Maldacena proposal \cite{hartmanmaldacena}; and it further supports our speculation that there is an effective low-energy notion of complexity that the action is calculating. We first discuss the tensor network picture for one side of the two-sided black hole; then we discuss the tensor network for the two-sided black hole and  examine its time development.\footnote{For additional work on the relationship between tensor networks and geometry, see also \cite{entrenholo,Swingle:2012wq,vidal_tns_geo,Pastawski:2015qua,Czech:2015qta,Czech:2015xna,Yang:2015uoa,Casini:2015zua}.}

Recall that a CFT is expected to have no long range entanglement on scales longer than the inverse temperature $1/T$. The basic picture is that the ground state renormalization group (RG) circuit remains a good description of the finite temperature RG circuit on scales shorter than $1/T$; at longer scales the state becomes trivial (unentangled). The RG circuit will typically differ from the ground state circuit as the scale $1/T$ is approached, but for simplicity we model the physics using the ground state circuit and a sharp cutoff at scale $1/T$.

The finite temperature state of the CFT is then
\beq
\rho(L,T) = V^k \rho(L/2^k,\Lambda_0)\( V^\dagger\)^k
\eeq
where $V$ is one circuit layer implementing coarse graining by a factor of $2$, $\Lambda_0$ is some UV energy scale [$\rho(L',T=\Lambda_0)$  has only cutoff-scale correlations], and $k = \log_2(\Lambda_0/T)$ is the RG depth. The system size is denoted by the argument $L$ in $\rho(L,T)$ and the number of lattice sites is $(L\Lambda_0)^{D-2}$ (recall that the CFT lives in $D-1$ spacetime dimensions; $D$ is the bulk spacetime dimension).

That was the description of one side of the black hole, but what about the two-sided black hole that purifies $\rho(L,T)$? Let the thermofield double state which purifies $\rho(L,T)$ be $\ket{\text{TFD}(L,T)}$. The RG circuit for the thermofield double state may then be written
\beq
\ket{\text{TFD}(L,T)} = V^k \otimes \(V^*\)^k \ket{\text{TFD}(L/2^k,\Lambda_0)}
\eeq
where $V^*$ appears because $\rho(L,T)$ and $\ket{\text{TFD}(L,T)}$ are related by a partial transpose on the second factor. The state $\ket{\text{TFD}(L',\Lambda_0)}$ roughly consists of $ (L' \Lambda_0)^{D-2}$ EPR pairs shared between the two sides of the thermofield double.

(In fact, this cannot be quite right because the spectrum of $\rho(L,T)$ is not completely degenerate---the thermofield double state is not perfectly maximally entangled---so some spread of the eigenvalues of the reduced state is necessary. However this is not an essential part of the story.)

\begin{figure}
\centering
\includegraphics[width=.7\textwidth]{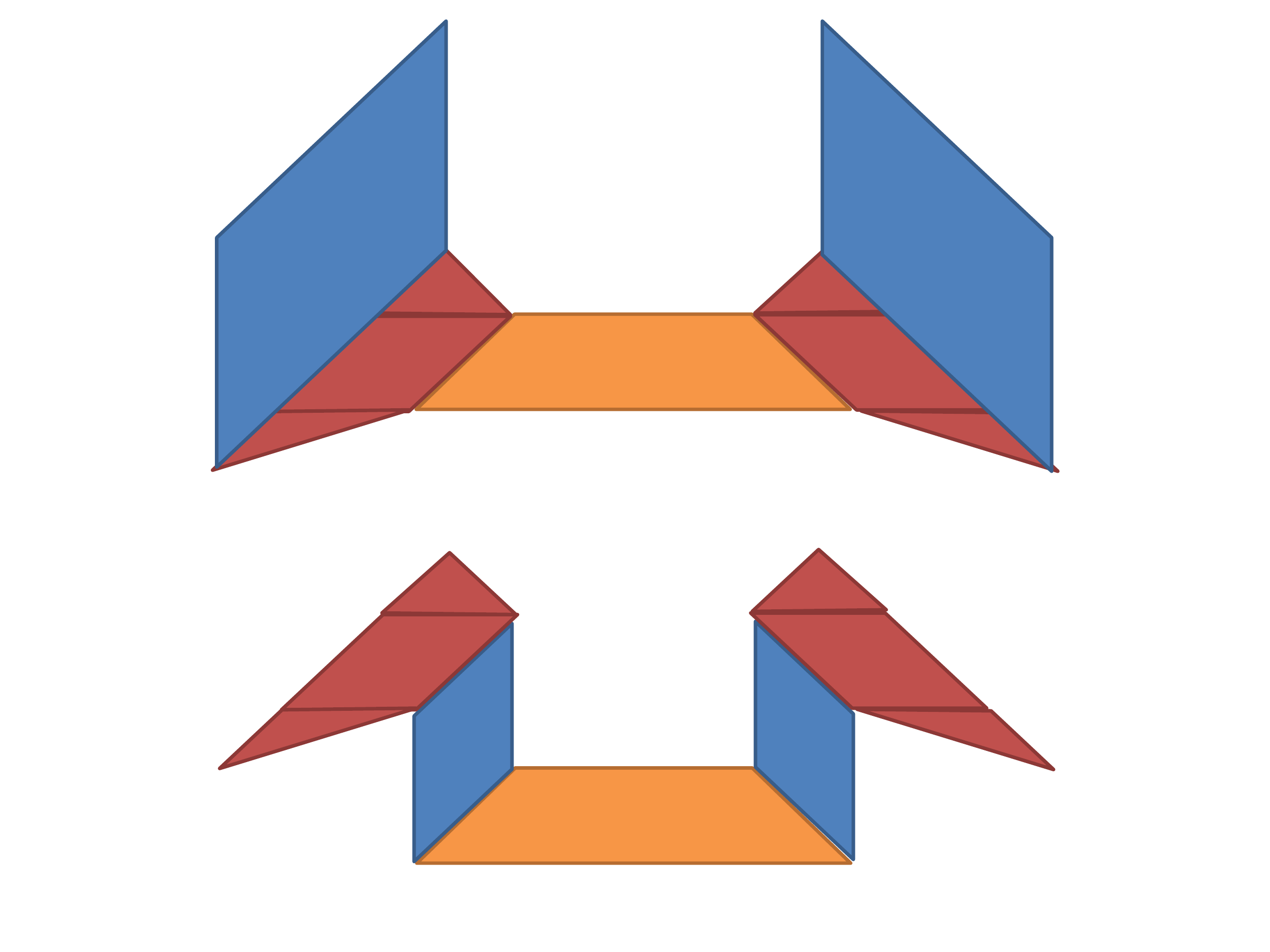}
\caption{Step 1: the time evolution acting on the UV degrees of freedom is renormalized into a time evolution acting on the IR degrees of freedom. The orange connection is the Einstein-Rosen bridge, the red region is the RG network, and the blue parts represent time evolution.}
\label{fig:HM1}
\end{figure}

\begin{figure}
\centering
\includegraphics[width=.7\textwidth]{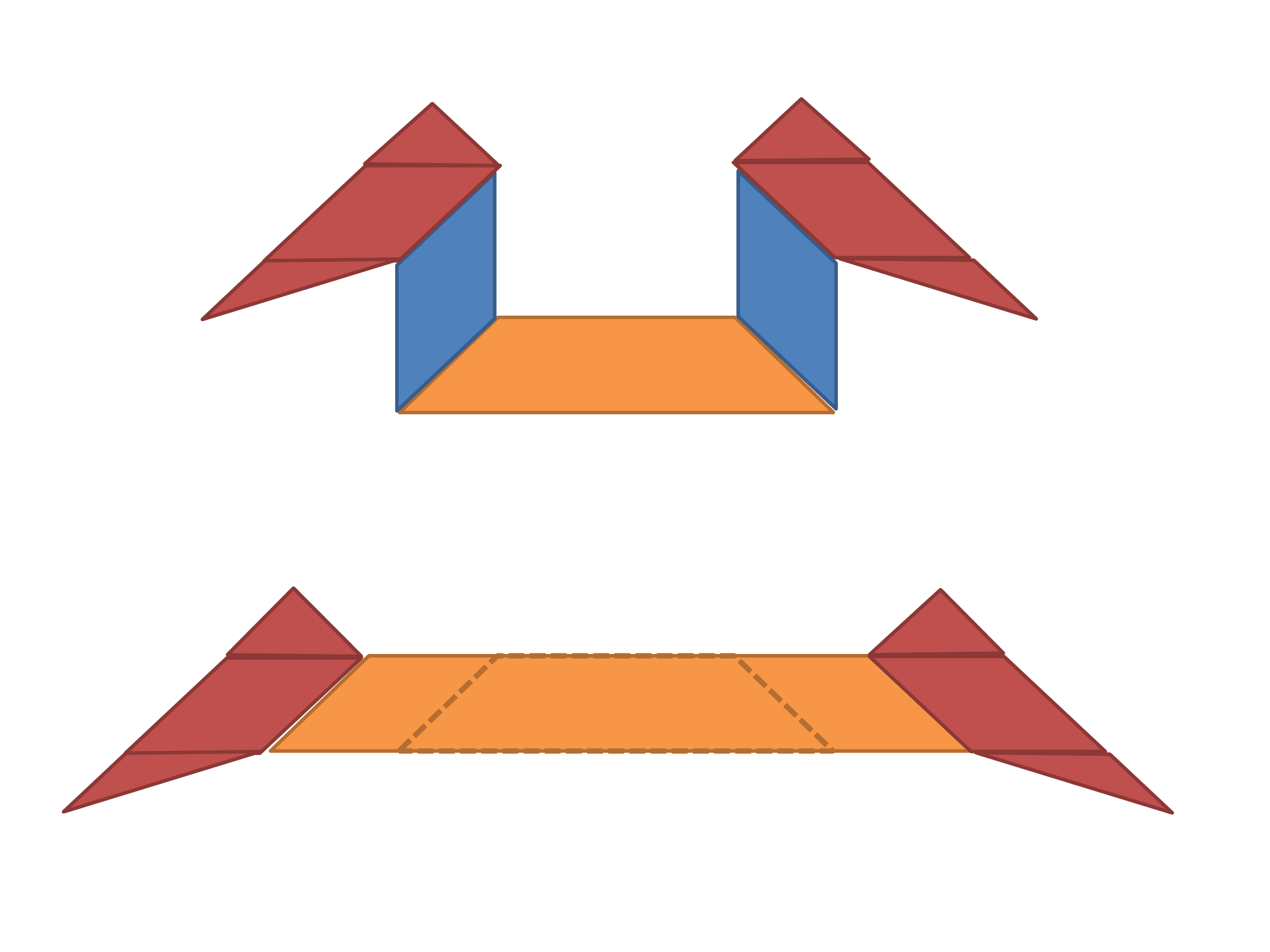}
\caption{Step 2: the IR time evolution is ``rotated" from vertical to horizontal and becomes a part of the Einstein-Rosen bridge. The orange connection is the Einstein-Rosen bridge, the red regions are the RG networks, and the blue parts represent time evolution.}
\label{fig:HM2}
\end{figure}

Now we time evolve the TFD state with the Hamiltonian $H \otimes I + I \otimes H$. This can be represented by a tensor network where we act on the  thermofield double state with $U(t)\otimes U(t)$, where $U(t) = e^{- i  Ht}$ is time evolution with respect to the UV Hamiltonian $H$. Let us ask about the complexity of the resulting state. A naive estimate is \beq
\CC_{\text{naive}} = \CC_{\text{RG}} + c  \Lambda_0 (L\Lambda_0)^{D-2} t,
\eeq
but in fact this UV divergent expression is only an upper bound. A much more efficient circuit can be found by first renormalizing the operator $U(t)$, see Figs. \ref{fig:HM1} and \ref{fig:HM2}.

To compute this renormalization it is useful to work with an infinitesimal time step: $U(\delta t) = I - i (\delta t)H$. The crucial piece of physics is that $H$ is a scaling operator with dimension $\Delta_H = 1$ (because it is the integral of the energy density which has dimension $D-1$). Hence we have
\beq
H^{(L)} V = V 2^{-\Delta_H} H^{(L/2)},
\eeq
where $V$ is one layer of the RG circuit,\footnote{This equation can in principle contain another term, call it $\CO$, on the RHS. This term must have the property that $V^\dagger \CO = 0$ and represents a high-energy contribution not captured by acting with $H^{(L/2)}$ and then $V$. It can nevertheless be argued to be irrelevant for the complexity growth. To fully show that the effective IR time evolution is approximately unitary one would have to examine this correction term more carefully.} and the parenthetical superscript indicates the size of the system. By definition we have $I^{(L)} V = V I^{(L/2)}$. It follows that for one RG step
\beq
\(I^{(L)} -i \delta t H^{(L)} \) V = V \(I^{(L/2)} - i \delta t 2^{-\Delta_H} H^{(L/2)} \)
\eeq
and that for $k$ RG steps
\beq
\(I^{(L)} -i \delta t H^{(L)}\) V^k = V^k \(I^{(L/2^k)} - i \delta t 2^{-\Delta_H k} H^{(L/2^k)}\).
\eeq
Thus time evolution at the UV scale for time $t$ can be traded for time evolution at a longer scale for time $ 2^{-\Delta_H k}t$. Alternatively, if the UV Hamiltonian $H_L$ has microscopic energy scale $\Lambda_0$ and acts on $(L\Lambda_0)^{D-2}$ sites, we may obtain the same action from a renormalized Hamiltonian with UV scale $\Lambda_0/2^k$ acting on $(L\Lambda_0/2^k)^{D-2}$ sites.

If the number of RG steps is $k= \log_2(\Lambda_0/T)$ then the scale of the renormalized Hamiltonian is $\Lambda_0/2^k = T$ and the number of sites on which it acts is $(L\Lambda_0/2^k)^{D-2} = (LT)^{D-2}$. Evolving for total time $t$, the complexity is now upper bounded by
\beq
\CC \leq \CC_{\text{RG}} + c \, T (LT)^{D-2} t \sim \CC_{\text{RG}} +   T S(T) t.
\eeq
This is illustrated in Fig.~\ref{fig:HM1}. The tensor network analysis therefore gives qualitatively the same result as we independently derived using CA-duality.

\subsubsection*{Complexity per gate is $O(c)$}

Implicit in the above analysis was the assumption that the complexity per gate in the tensor network is roughly $c$ for a CFT$_2$ (with the obvious generalization to higher dimensions). This is not a trivial assertion; for example, it was proposed \cite{Swingle:2012wq} that random tensors make a good starting point for thinking about the entanglement properties of holographic field theories. However, with high probability a random tensor would be of at least scrambling complexity $O(c \log c)$ and possibly (depending on the ensemble from which the tensor is drawn) of much higher complexity. Recently random tensors have been further advanced \cite{Yang:2015uoa,RandomModelHolgraphy,Hosur:2015ylk} as toy models of holography and have been shown to satisfactorily model the entanglement structure of the field theory in terms of bulk minimal surfaces. However here we argue that such models do not correctly capture the complexity properties of the holographic field theory.

Let us justify this claim. First, note that each line in the tensor network should be understood as consisting of $O(c)$ degrees of freedom bundled together. In other words, every site of the lattice theory has $O(c)$ degrees of freedom, so if, for example, the field theory consists of a large-$N$ SU$(N)$ gauge theory, then each lattice site consists of a large-$N$ matrix quantum mechanics (coupled to the other sites) and we expect $c \sim N^2$.

It is easiest to first analyze the part of the network corresponding to conventional time evolution (the wormhole part). Implementing one layer of gates in the circuit corresponds to evolving with our properly normalized local Hamiltonian for a time of order the inverse temperature. Since the local Hamiltonian acts on $O(c)$ degrees of freedom per site and we evolve for a time of order $\beta$, the complexity of the resulting tensor is also $O(c)$. This implies that the rate of increase in complexity is proportional to the product of the central charge $c$, the number of thermal cells $(L T)^{D-2}$, and the temperature $T$. This combines to give $ST$.

Now let us analyze the RG part of the network (the part outside the horizon). Suppose we are performing a $2\rightarrow 1$ RG procedure in one boundary spatial dimension ($D=3$) by mapping the ground state on $L$ sites to the ground state on $L/2$ sites. This mapping is accomplished by some unitary transformation acting on the local degrees of freedom. Existence of a smooth continuum limit suggests that this unitary is the exponential of a local Hermitian generator. In other words, the RG generator should share the rough characteristics of an ordinary local Hamiltonian acting on the system.

The basic point is then that to achieve scrambling complexity   $O(c \log c)$ for the local gates, we would have to run the local evolution for a time $O(\log c)$. However, because the local generator also couples neighboring sites, correlations will be generated at scale $O(\log c)$. This is unphysical---correlations of $O(\log c)$ are not compatible with bulk locality on the AdS scale. Hence the RG transformation must correspond to running a local generator for a RG time of order one. This implies that the complexity of the gates in the RG transformation are $O(c)$ as claimed.

\section{Discussion}
\label{sec:discussion}

There is no doubt that the interior of a black hole grows with time until some kind of nonperturbative quantum effect saturates the growth, probably at an exponential time. Moreover there is no doubt that the complexity of the dual gauge theory state grows long after it has relaxed to thermal equilibrium. Identifying these growth phenomena is the basis for the complexity-geometry duality. In this paper we have proposed a new form of this duality that eliminates some of the less attractive aspects of CV-duality. The resulting CA-duality says that
 the complexity of a boundary state is dual to the action of the corresponding Wheeler-DeWitt patch in the bulk.

 In this section we will consider some of the implications of CA-duality as well as a sample of open questions for future work.

\subsection{Comparing CA-duality and CV-duality}
\label{subsec:implications}

From the perspective of CA-duality, CV-duality \cite{complexityshocks} can be expressed as follows:
\begin{itemize}
\item[1)] The geometry of an Einstein-Rosen bridge may be identified with maximal (spatial) volume slices.

\item[2)] The rate of complexity increase is bounded by the product of entropy and temperature\footnote{Reference \cite{complexityshocks} did not claim that $TS$ was a bound on complexity growth; only that it was a natural expectation for the rate of growth of complexity for a black hole.}
\be
\frac{d\CC}{dt}  \sim TS.
\ee
\item[3)] Black holes saturate the bound.
\end{itemize}

CA-duality sharpens assumption \#3. Assumptions \#1 and \#2 are different, but for many black holes the  length scale in CV-duality may be chosen so that the predictions are roughly the same.  We have already discussed the fact that an Einstein-Rosen bridge is a long tubelike geometry. The relation between the spatial volume of the tube (defined on the maximal slice) and the spacetime volume $|\CW|$ of the Wheeler-DeWitt patch has the form (for a large AdS-Schwarzschild black hole)
\be
|\CW|  \sim V \lads.
\ee
Combining the various ingredients described in \S\ref{subsec:CA} one finds that the form of the earlier proposal could be written as ``complexity $\sim$ action.''

Assumption \#2 is also closely related to the present proposal. In all cases that we have studied the product $TS$ is within a factor of a few of the rate of change of action. For example, for neutral black holes we found the rate of change of action is proportional to the mass of the black hole. The product $TS$ and the mass are related by
\bea
TS &=& \bigg( \frac{D}{D-1} \bigg)M  \ \ \ \ \ \textrm{(large black holes}, r_h \gg \lads)    \cr \cr
TS &=& \bigg( \frac{D-2}{D-1} \bigg)M  \ \ \ \ \  \textrm{(small black holes}, r_h \ll \lads) .\
\eea

For charged and rotating  black holes, the mass itself is not a good measure of complexity growth, but the rate of change of action is still within a factor of a few of $TS.$
Thus, at least for the special solutions we have considered, the quantitative implications of CA-duality are not very different from those of CV-duality.

However CA-duality  provides a degree of universality that CV-duality lacked. The same universal constant connects action and complexity for all neutral black holes that we have studied, and in each case the computed complexity saturates the appropriate bound.

\subsection{Open questions}

A number of open questions remain.
\subsubsection{The definition of complexity}

In this paper we have defined complexity as the number of primitive gates in the minimal quantum circuit that builds our state. This definition is not ideal. For example, in our definition the value of the complexity depends on the choice of  the set of primitive gates.

One open question is where there is a better definition. In particular, is there a definition that is suitable for continuous Hamiltonian systems and that can give meaning to the prefactor in the bound Eq.~\ref{C-bound}?

Further, assuming such a definition can be found, is the universality of our results indicating a universal dynamics of black holes, independent of mass, dimension, rotation, etc?

For an initial attempt to address these questions, see Appendix~\ref{sec:appendix:A1}.

\subsubsection{CA-duality  as a diagnostic tool}

Assuming CA-duality, small Reissner-Nordstrom charged black holes saturate our bound on complexification, and large (compared to the AdS radius) charged black  holes violate the naive bound. But in the case of large charged black holes, we understand what goes wrong---the problem is not with our conjecture, the problem is that large charged black holes are not described by the hairless RN solution. The RN truncation does not fully capture the essential physics of UV-complete holographic theories, and the hair renders our calculations unreliable.

The difficulties of UV-completing large RN black holes was already known. In situations that are not so well studied, could the CA-duality be used to diagnose an inconsistent truncation?

\subsubsection{Diagnosing transparency of horizons}

One interesting relation between complexity and Einstein-Rosen bridges involves the ``transparency" or ``opacity" of the horizon: it has been conjectured that black holes with growing complexity have transparent horizons \cite{opacity} while those with decreasing complexity have opaque horizons. This result found support in the shock wave analysis, both for CV-duality (see \cite{opacity}) and, in this paper, for CA-duality.

Although the results are very similar for CA-duality and CV-duality, calculations are generally much simpler for CA-duality when shock waves are involved. No differential equation needs to be solved in order to locate the maximal slice. CA-duality only requires integrals to be computed, and these are usually elementary. Thus diagnosing the  transparency of event horizons \cite{opacity} is much simpler using CA-duality than CV-duality.

\subsubsection{Weak coupling and stringy corrections}
In this paper, we have worked exclusively with strongly coupled holographic theories described in the bulk by Einstein gravity. It is natural to ask how our results and conjecture should be affected by including stringy corrections in the bulk.
In the case of $\CN=4$ $U(N)$ SYM in $D-1=4$ where the 't Hooft coupling $\lambda$ is related to the string scale $\ell_{s}$ by $\lambda = (\lads/\ell_{s})^4$, the vanishing string scale implies infinite field theory coupling.

First, we will remind the reader of the results in a related holographic information processing scenario: scrambling and chaos. Black holes (described by Einstein gravity) are the fastest scramblers in nature \cite{bhmirror,fastscramble,Maldacena:2015waa}, meaning they're the fastest at mixing their quantum information. In \cite{Shenker:2014cwa}, it was shown that stringy corrections to the butterfly effect increase the black hole scrambling time. At zero coupling, the theory is free and does not scramble at all.

In many respects, the principles of scrambling and complexity growth are closely tied together. At a qualitative level, if a system cannot  mix together information from across  its degrees of freedom, it cannot compute \cite{Hosur:2015ylk}.\footnote{In some cases, we can quantitatively think of complexity as the integral of the ``size'' of an operator over time \cite{localshocks}, where the growth of the operator is a manifestation of the system scrambling. An example of this is given by the operator shown in Fig.~\ref{fig-localized-tensor-network}.} Thus, we expect that stringy corrections should reduce the computation rate of black hole solutions, pulling them away from saturating the conjectured bound on complexity growth \eqref{eq:LloydBound}.

Weakly coupled systems compute slowly because collisions are rare. Free systems barely compute at all because there are no collisions. It would be interesting to compute the complexification rate at weak coupling or in the presence of stringy corrections.

\subsubsection{Regularizing the action} \label{sec:corner}

In this paper we have calculated the rate of growth of action of the WDW patch at late times. If we wish to calculate the rate of growth at finite times we must introduce a method of regularizing divergences, for example those that occur at the boundary of AdS. A divergence of this type shows up in the part of the WDW patch that lies behind the past horizon. The corner term at the bottom, which consists of the intersection of two lights heets, is divergent.\footnote{We thank Henry Maxfield for helpful discussions that prompted us to revisit this issue.}

Let us take for example the BTZ black hole. One way of regulating the action using timelike radial geodesics leads to the formula
\begin{equation}
\begin{aligned}
\frac{d\CA}{dt}=\frac{r_h^2}{4G\lads^2}\left(\tanh^2(\pi t/\beta)+\frac{\log\left(\cosh(\pi t/\beta)\right)}{\cosh^2(\pi t/\beta)} - \frac{\log\epsilon}{\cosh^2(\pi t/\beta)}\right)+\mathcal{O}(\epsilon),
\end{aligned}
\end{equation}
where $t \equiv t_L + t_R$ and $\epsilon$ is an ultraviolet regulator. There are three terms in parentheses.
The first term  tends to the constant late-time growth rate from Sec.~\ref{sec:actionuncharged}. This correctly gives the linear growth that has been the subject of this paper and is robust against changes in the cutoff prescription. The second term is a UV-finite transient, which dies away exponentially fast. The third term is more problematic. While it exponentially decreases with time, it is logarithmically UV divergent.

A UV-divergent complexification rate violates the bound of Eq.~\ref{C-bound}, and mixes IR and UV dependence in an anomalous way. We do not understand the physics of this term. It seems likely there is an improved regulator that removes the divergence. This UV divergence is  tied up with the issue of the complexity of formation (the complexity already present at $t_L=t_R=0$) and we intend to address  this question in \cite{ToAppear}.

Because there is no corner, this issue does not arise for either one-sided black holes or geometries with a wide Penrose diagram, e.g. wormholes lengthened by multiple shock waves. So while a subtlety with the regulator does occur for the action growth of unperturbed eternal black holes, these black holes are dual to rather special states of the conformal field theory. For many other states, such as those dual to black holes formed by collapse or wormholes perturbed by multiple strong shock waves, CA-duality gives a robust regulator-independent characterization of the growth of complexity.

\subsubsection{Action from near the singularity}
In  CV-duality, the maximal slice stayed safely away from the black hole singularity. In  CA-duality, for the Schwarzschild-AdS black hole the WDW patch extends all the way to $r=0$.

Near the singularity, semiclassical physics breaks down. For the black holes have considered in this paper, this is only a small concern because the action contribution from near the singularity is small: the result is insensitive to precisely how close the cutoff is to $r=0$. However, in more complicated situations it is conceivable that the action could be sensitive to the location of the cutoff near the singularity. It would be interesting to explore this question.

\subsubsection{Boundary terms}
The classical equations of motion do not uniquely define the action. For example, adding an overall constant to the action has no effect. More generally, any function of just the intrinsic properties of the boundary will be ignored by the variational principle and leave the equations of motion unaffected.

This ambiguity is particularly relevant to our conjecture, since we assign a meaning to the numerical value of the action, and since our boundaries are not only out near the asymptotic edge of AdS, but penetrate deep into the bulk, sometimes all the way to the singularity.

This is not just a hypothetical concern. Electric-magnetic duality in four dimensions suggests that magnetically charged black holes should complexify at the same rate as their electrically charged duals. The bulk Maxwell contribution to the action, however, changes sign
\begin{equation}
F_{\mu \nu} F^{\mu \nu} \sim  \vec{B}^2 -\vec{E}^2.
\end{equation}
Electric and magnetic black holes can be returned to an equal footing by adding a boundary term that is sensitive to the magnetic field near infinity \cite{Braden:1990hw,Hawking:1995ap}. The question is, what is the principle that instructs us which boundary terms to include and which to omit?

There are a number of equivalent ways to phrase this question \cite{Braden:1990hw,Hawking:1995ap}. Instead of asking what boundary terms to include in the action, we could  ask what quantities we intend to hold fixed at the boundary (for example electric charge or  electric potential), or ask what ensemble our system is in, or which free energy the Euclidean version of the action is to calculate.

Our prescription is that  there should be no contribution to the action from the intrinsic geometry of the boundary. The extrinsic terms on the boundary are determined by the demand that all interior boundaries be permeable, in the sense that conserved charges can pass through them (so that for example it is the electric potential and not the charge that is fixed). It is reassuring that by adopting this principle the $Q \rightarrow 0$ limit of the charged black holes case, Eq.~\ref{eq:chargedactionincrement}, reduced to the result for the uncharged $Q=0$ answer Eq.~\ref{eq:unchargedactionincrement}; had we adopted any other boundary terms, the $Q \rightarrow 0$ limit would have been discontinuous.

\subsubsection{Theories without action formulations}

Our conjecture assigns a physical meaning to the value of the action.  It would not be clear what to do if the bulk theory did not have an action formulation, or if the action wasn't a real number,
or if the bulk theory  had more than one action formulation. These are unlikely to make contributions at leading order in $N$, but it would be interesting to explore these issues further.


\subsubsection{Very early times}

Throughout this paper we have generally been considering the rate of complexification at times that are neither too early nor too late.

At early times, the rate of the increase of action is slow. For example, for the uncharged black hole in $D > 3$, $d\CA/dt$ is zero when $t_R = t_L = 0$, and remains zero up until the past light sheets first cross; it then grows,  asymptoting to the answer quoted in Eq.~\ref{eq:unchargedactionincrement}.

\begin{figure}[htbp] 
   \centering
   \includegraphics[width=5in]{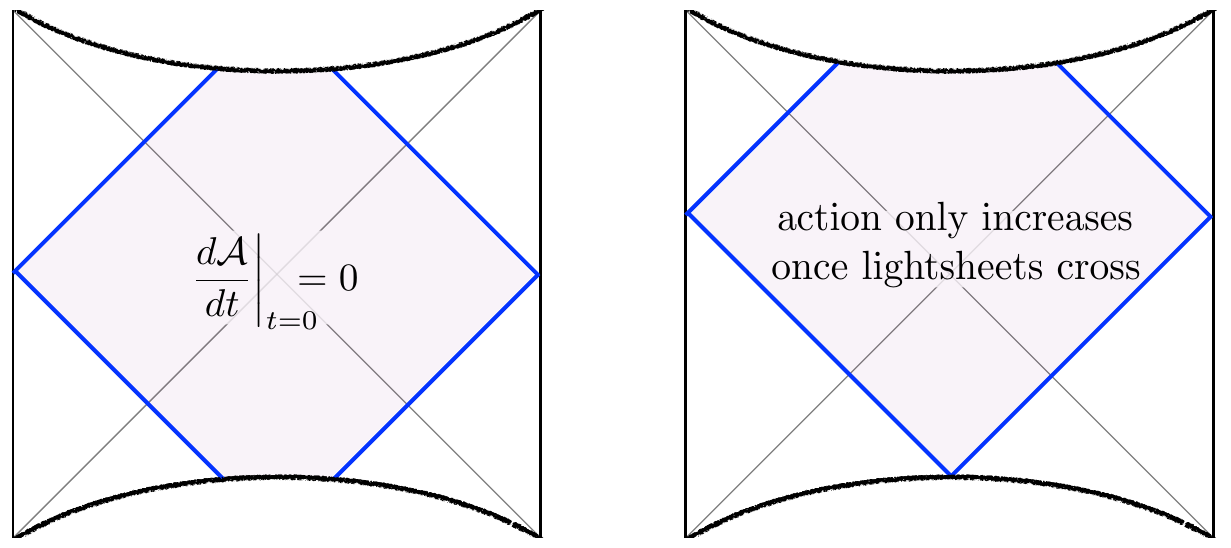}
   \caption{At early times, the action of the WDW patch of a neutral AdS black hole does not change with time. The action starts to increase only after enough time has passed that the light sheets cross.  The critical moment at which the action starts to increase is shown on the right.}
   \label{fig:earlytimes}
\end{figure}

So long as the ingoing light sheets don't meet before they hit the singularity, there are two separate time translation symmetries: a $t_L$ symmetry and a $t_R$ symmetry. Once the two light sheets meet, these two symmetries break to a single overall time translation symmetry in $t_L + t_R$. It is only then that the action can start growing. This is shown in Fig.~\ref{fig:earlytimes}. (This fact is particularly obvious from the perspective of the ``past wedge'' calculation of Appendix~\ref{sec:pastwedge}.)

The period of constant action  exists because the singularity of an AdS black hole bows in for $D \geq4$ \cite{Fidkowski:2003nf}; for small black holes the action begins to increase after about an AdS time (which can be very long compared to the Schwarzschild time), and for large black holes it increases after about a thermal time. (This effect is absent for one-sided black holes that form from the collapse of a null shell---in that case the complexity starts to increase the moment the shell is emitted from the boundary.)

If CA-duality can be trusted on time scales this short, it would be interesting to develop a CFT understanding of why the holographic dual does not immediately begin complexifying. For large black holes, this would involve understanding the reference state relative to which the computational complexity is defined. At late times it suffices to consider the reference state to be the thermofield double state, but the deviation from linear growth at early times suggests that there may be a more primitive reference state from which the thermofield double state and the $(t_L + t_R)=\lads/2$ state will be the same complexity distance. (We will return to this issue in \cite{ToAppear}.)

For small black holes it may be harder to understand the early time behavior because even outside the context of complexity small black holes are not holographically well understood.

\subsubsection{Very late times}

At very late times, of order the classical recurrence time $e^S$, the complexity must saturate and stop increasing \cite{opacity}. It is a nontrivial test of our duality that there is an instanton that invalidates the semiclassical bulk description after a similar time scale. After a time of order $e^S$ it is likely that at some stage the large black hole will have undergone a thermal fluctuation down to the size of a small AdS black hole (plus thermal gas), which would then evaporate (and then recollapse and reform the black hole).
This signals a breakdown of the semiclassical spacetime description, because the huge entanglement is carried not by the semiclassical spacetime but by the decay products \cite{Maldacena:2001kr,Maldacena:2013xja}.  It would be interesting to investigate if there are other properties of these two dual processes that can be connected.

At extremely late times, of order the quantum recurrence time $e^{e^S}$, the complexity will undergo its first recurrence and become temporarily small again. It is not clear that the semiclassical bulk description has anything to say about times this late.

\subsubsection{Principle of least computation}

From the perspective of tensor networks, the course-grained geometry encodes the minimal quantum circuit that prepares the state. In this paper we have argued that the complexity of the boundary state is given by the action of the bulk. In the bulk, the classical equations of motion are given by the principle of least action. Can gravity be understood via a principle of least computation, and if so what is the  quantum generalization?

\section*{Acknowledgments}
We thank Patrick Hayden, Don Marolf,  Henry Maxfield, Simon Ross, Jorge Santos, and Douglas Stanford for useful discussions. AB thanks the participants at Perimeter Institute's ``Quantum Information in Quantum Gravity II'' conference for useful feedback.  We thank Douglas Stanford for his generous donation of three figures from previous works.

DR is supported by the Fannie and John Hertz Foundation and is very thankful for the hospitality of the Stanford Institute for Theoretical Physics during various stages of this work. DR also acknowledges the U.S. Department of Energy under cooperative research agreement Contract No.~DE-SC00012567. BS is supported by the Simons Foundation. This work was supported in part by National Science Foundation Grant No.~0756174 and by a grant from the John Templeton Foundation.
The opinions expressed in this publication are those of the authors and do not necessarily
reflect the views of the funding agencies.

\appendix

\numberwithin{equation}{section}

\section{Further comments on the complexification bound}
\label{sec:boundapp}

If we want a notion of complexity such that local Hamiltonian evolution leads to complexity growth linear in $t$ and the number $n$ of active degrees of freedom, then the Hamiltonian should first be normalized so that the energy is extensive in $n$. The complexity can then be preliminarily defined by a limiting procedure where we take the gate set $\{G_\alpha\}$ to consist of elements which are close to the identity, $G_\alpha = I + i \delta H_\alpha$, with the $H_\alpha$ taken to be simple Hermitian operators. As $\delta \rightarrow 0$ the number of gates diverges as $1/\delta$, so the complexity may be defined as
\beq
\CC \propto \lim_{\delta \rightarrow 0} \delta N_{\text{gates}}.
\eeq

The above definition is roughly equivalent to saying the complexity is defined as an $L_1$-norm (with a high penalty for nonsimple directions) of a tangent vector to the manifold U$(2^n)$ of many-body unitaries. Two comments are necessary. First, the limiting procedure is important because with a fixed noninfinitesimal gate set the number of gates grows like $\frac{t \log\(\frac{t}{\epsilon}\)}{\log\(\log\(\frac{t}{\epsilon}\)\)}$, i.e. not quite linearly in $t$ \cite{opthamsim}. Second, our definition is inspired by Nielsen's complexity geometry \cite{nielsenmetric} where the complexity grows linearly in $t$. However, \cite{nielsenmetric} uses an $L_2$-norm so the complexity growth is proportional to $\sqrt{n}$ instead of $n$ in that formulation.

We might further hope that the necessary notion of approximation and other details in the definition of the complexity can be swept up into the overall prefactor which is not fixed by the considerations in this appendix. In the main text we assumed that the prefactor could be chosen so that the stated complexity growth bound holds. While not proven, this seems reasonable given that there are special information theoretic properties \cite{infocurvelength} (such as concentration of eigenvalues) of field theories with semiclassical holographic duals.

A rough argument for the complexification bound proceeds as follows. Decompose $e^{-i H t}$ as
\beq
e^{-i H t} = U_H e^{-i h t} U_H^\dagger
\eeq
where $U_H$ is a unitary which maps from the energy basis to a tensor-product basis and $h$ is a diagonal matrix whose entries are the energy eigenvalues. Provided we only care about energies up to roughly $E = \langle\psi| H|\psi \rangle$ one can argue using adiabatic evolution that the complexity of $U_H$ is of order $e^S$ where $S$ is the microcanonical entropy at energy $E$. From this perspective we may conceptualize the action of $e^{-iHt}$ as a dephasing process in which the $e^S$ complexity of $U_H$ is slowly revealed as the diagonal term $e^{-i h t}$ increasingly inhibits the cancellation of $U_H$ and $U_H^\dagger$. Assuming the input state has maximum energy $E$ we would again expect the rate of ``increasing failure to cancel" to be bounded by $E$.

The above argument sketch has at least one major problem: it is violated by ``cat states", that is by states which are superpositions of macroscopically different states. The simplest example contains a noncomputing branch and a rapidly computing branch,
\beq
\ket{\theta } = \cos \theta \ket{0} + \sin \theta \ket{E},
\eeq
where $\ket{0}$ is the ground state and $\ket{E}$ is a highly excited state. If $\theta \ll 1$ then the average energy of $\ket{\theta}$ is approximately $\sin^2 \theta E \ll E$ but the rate of complexification of $\ket{\theta}$ is likely given by $E$ (without any $\sin^2 \theta$ factor). This is an arbitrarily bad violation of the complexity growth bound.

Another very serious objection comes from considering superpositions of multiple computers. Suppose the system is a composite of $i=1,\dots,m$ computing systems and suppose the Hamiltonian is $H = \sum_i H_i$ where each $H_i$ acts only on the $i$th computer. Then a state of the form
\beq
\ket{\text{cat}} = \frac{1}{\sqrt{m}} \sum_{i=1}^m \ket{0}_1 \dots \ket{E}_i \dots \ket{0}_m
\eeq
has average energy $E$ but likely complexifies at a rate $m E$. Again, we appear to have an arbitrarily bad violation of the complexity growth bound.

However, both of these states correspond, on the gravity side, to superpositions of black holes and ground states. Such energy cat states are certainly not semiclassical, so it remains possible that among semiclassical states the complexity growth bound could be obeyed. As for more general quantum computers, bearing in mind the ``cat computer" examples, it is at present unclear to us if there is any general bound on the rate of computation.

\subsection{Comments on Hamiltonian locality and gate simplicity} \label{sec:appendix:A1}

In the previous section, it was mentioned that the elements of the gate set  $\{G_\alpha\}$ are taken to be close to the identity $G_\alpha = I + i \delta H_\alpha$, with the $H_\alpha$ taken to be simple Hermitian operators. Here, we will elaborate a bit on what is meant by simple in this context, and whether it is reasonable to have a bound on complexity.\footnote{We thank Douglas Stanford for questions and discussions leading to the creation of this section.}

An individual $H_\alpha$ will be a product of operators  at neighboring points, with the number of such operators defining the size or weight of the $H_\alpha$. We will call a gate set $k$-local if all elements of $\{H_\alpha\}$ have a size less than or equal to $k$. When we say that the $H_\alpha$ should be simple, what we mean is that $k$ should be small.

In this paper, we are mostly interested in estimating the complexity of Hamiltonian time evolution. Thus, our gate set should be adept at approximating $U(t)=e^{-iHt}$. In the limit of small $\delta$, we can suggestively write a gate as $G_\alpha = e^{i \delta H_\alpha }$. This suggests a reasonable choice for the $\{H_\alpha\}$ could be the individual terms in the Hamiltonian. Certainly this makes it clear how this gate set can approximate  Hamiltonian time evolution.

Even with a more general gate set, this also suggests a reasonable value for $k$. If the Hamiltonian is $k$-local, then we should at least choose a gate set that is also only $k$-local. There's some reasonable intuition behind this: a Hamiltonian with locality $k$ means that interactions are roughly spread over at most $k$ degrees of freedom. Presumably in a circuit approximation of Hamiltonian evolution, one should only be able to use gates that mix together at most $k$ degrees of freedom per gate. However, by measuring the complexity with a gate set that either depends directly on the $H$ (by explicitly choosing the $H_\alpha$ to be the terms in $H$) or by picking the $k$-locality of the $H_\alpha$ to equal the $k$-locality of the Hamiltonian, we see that the complexity $\CC$ will be a function of both the Hamiltonian and the state. By changing the properties of $H$, we have to change the way in which we measure $\CC$.\footnote{In particular, we thank Douglas Stanford for emphasizing this point.}

Instead, let's consider the more general case, where we choose a $j$-local gate set, but have a $k$-local Hamiltonian. In that case, $\CC$ has the desirable property of being a function of $j$ and the state $|\psi \rangle$, but no longer a function of $H$. Now, it's easy to see that the bound discussed in this paper
\be
\frac{d\CC}{dt} \le \frac{2 }{\pi \hbar}\langle \psi | H |\psi \rangle,\label{eq:appendix-bound}
\ee
cannot hold without modification. As a simple example, consider the case of $k \gg j$. (For instance if the system has $N$ degrees of freedom, $H$ is a random $2^N \times 2^N$ Hermitian matrix, and $j$ is $O(1)$.) For $k \sim N$, a small amount of time evolution will lead to near exponential complexity as measured by the $j$-local gate set and violate the bound. Therefore, we need to consider the dependence on both $j$ and $k$ explicitly
\be
 \frac{d\CC}{dt} \le \frac{g(k)}{f(j)}\, \frac{2 }{\pi \hbar}\langle \psi | H |\psi \rangle ,\label{eq:appendix-bound-depends-on-locality}
\ee
where the monotonically increasing functions $f(j), g(k)$ capture the dependence of the complexity on the locality of the gate set and the Hamiltonian. A reasonable assumption is that these are the same functions of their argument (since they're trying to account for the same type of dependence). Thus, for the natural choice discussed above $j=k$, these factors cancel, and Eq.~\ref{eq:appendix-bound-depends-on-locality} reduces to Eq.~\ref{eq:appendix-bound}. For the case considered previously $k \gg j$, we see that the bound becomes harder and harder to violate, compensating for the large increase of complexity under time evolution as measured by the $j$-local gate set. In the other limit of $k \ll j$, the rate of complexity growth is vanishing. This is because the gate set is so large that the reference system is never more than $O(1)$ gates away from the time-evolved state; the complexity will always stay small.

\subsubsection*{A comment on the relevance of the ground state}
The complexification bound we propose has the unusual property of depending on the energy of the ground state of the Hamiltonian $H$. Given the ground state, $|0\rangle$, and a state of interest, $|\psi\rangle$, one might think that it could be possible to change the energy of the ground state $\langle 0 | H |0 \rangle$ without altering the complexodynamics of  $\langle \psi | H | \psi\rangle$. By arbitrarily lowering the ground state, it appears we could make the difference $\langle \psi | H | \psi\rangle - \langle 0 | H | 0 \rangle$ arbitrarily large.

Let's explore this idea a bit further. We can lower the ground state by an amount $\Delta > 0$ by adding a term to the Hamiltonian
\be
H' = H - \Delta | 0 \rangle\langle 0 |.
\ee
The ground state of the perturbed Hamiltonian $H'$ has ground state energy $\langle 0 | H | 0 \rangle - \Delta < \langle 0 | H | 0 \rangle$. Additionally, if the ground state has negligible support on $|\psi\rangle$ such that $\langle 0 |\psi\rangle \ll 1$, then the complexodynamics remain unchanged $\langle \psi | H' | \psi\rangle \approx \langle \psi | H | \psi\rangle$. This appears to let us  shift the complexification bound by an arbitrary amount. However, the catch is that the ground state projector $| 0 \rangle\langle 0 |$ is a high-weight operator of $O(N)$ and will destroy the locality of the Hamiltonian.

If instead we take $\Delta < 0$ and try to make the complexification bound tighter, the ground state will quickly cross the gap and become an excited state. Since the gap is $O(N^{-1})$, this would only be a $1/N$ correction to the bound.

\section{Spherical shock waves at finite time and energy} \label{appendix-finite-time}

\begin{figure}
\begin{center}
\includegraphics[scale=.7]{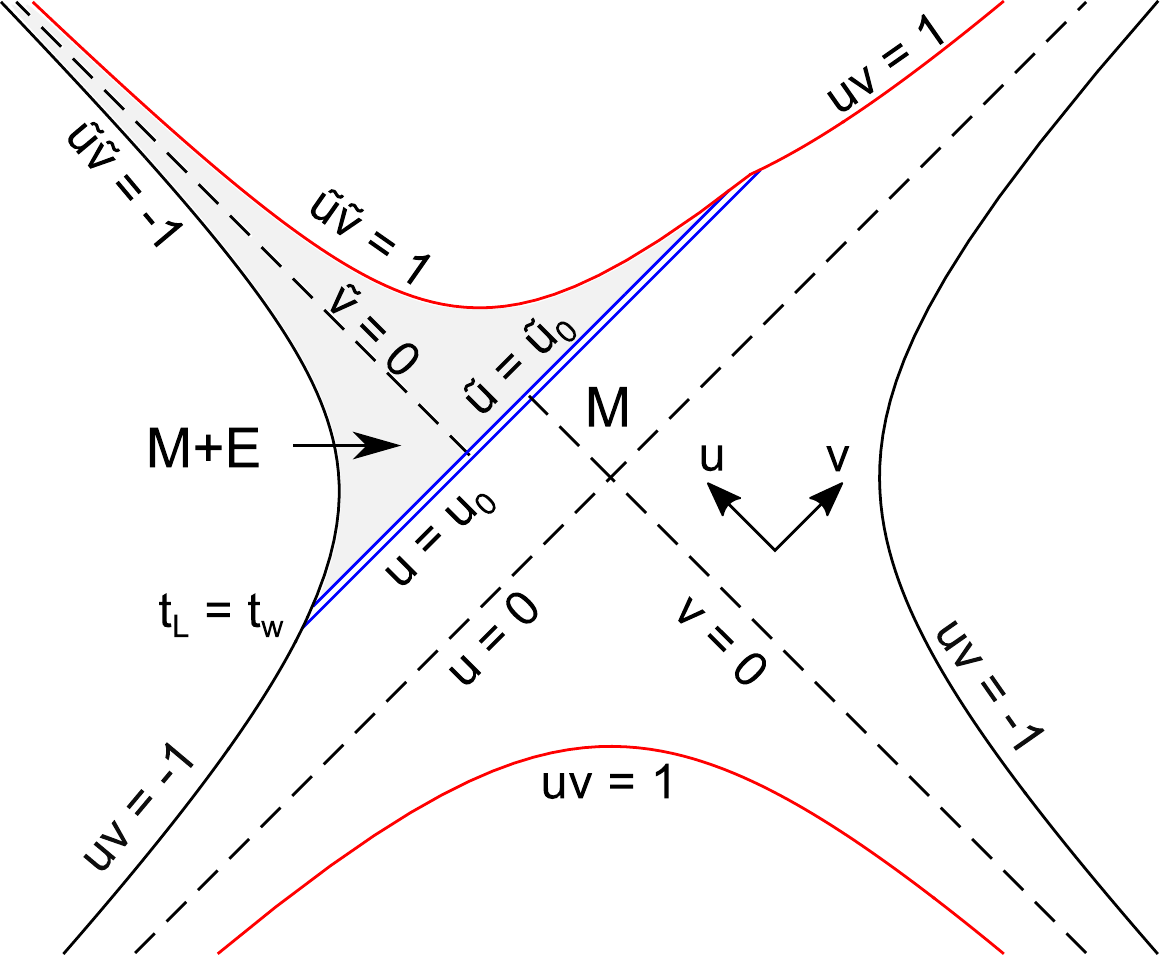}
\caption{Kruskal diagram of shock wave at finite time $t_w$. The white region to the right of the shock has energy $M$ and is covered by coordinates $(u,v)$. The gray region to the left of the shock has energy $M+E$ and is covered by coordinates $(\tilde{u}, \tilde{v})$. The shock is drawn in blue and travels along the surface $u_0 = e^{-r_{h} t_w / \lads^2}$ or equivalently $\tilde{u}_0 = e^{-\tilde{r}_{h} t_w / \lads^2}$.}\label{finite}
\end{center}
\end{figure}

 We will work in $D=2+1$ dimensions for simplicity. The shock wave has energy $E$ and the black hole has energy $M$, so we want to patch together a spacetime with energy $E+M$ (to the left of the shock) and a spacetime with energy $M$ (to the right of the shock). See Fig.~\ref{finite}. We will follow \cite{Dray:1985yt,shock} for the matching conditions between the two spacetimes.

Let us associate coordinates $(u,v)$ with the patch with energy $M$ and $(\tilde{u}, \tilde{v})$ with the patch with energy $E+M$. Because of the increase in mass, the horizons are at different radii in the two spacetimes, $r_{h}$ and $\tilde{r}_{h}$, respectively. In the patch with energy $M$, the relationship between Kruskal coordinates and Schwarzschild coordinates in the asymptotic region to the left of the horizon is
\be
u = \sqrt{\frac{r-r_{h}}{r+r_{h}}} \,e^{-r_{h} t_L / \lads^2},  \qquad v = -\sqrt{\frac{r-r_{h}}{r+r_{h}}}\, e^{r_{h} t_L / \lads^2}, \label{uv}
\ee
and in the patch with energy $M+E$ the relationship is\footnote{We have fixed a relative boost ambiguity by demanding that the boundary time $t_L$ is continuous.}
\be
\tilde{u} = \sqrt{\frac{r-\tilde{r}_{h}}{r+\tilde{r}_{h}}}\, e^{-\tilde{r}_{h} t_L / \lads^2},  \qquad \tilde{v} = -\sqrt{\frac{r-\tilde{r}_{h}}{r+\tilde{r}_{h}}}\, e^{\tilde{r}_{h} t_L / \lads^2}, \label{tilde}
\ee
where on this boundary, time runs backwards so that the upper left of Fig.~\ref{finite} is at $t_L = - \infty$.\footnote{Note: in this Appendix we have reversed our time conventions for precursors $W(t)$. Here, we will take both left boundary Hamiltonian time evolution $t_L$ and Killing time evolution in the bulk to increase to the past. } We must have $\tilde{r}_{h} > r_{h}$ since the horizon jumps out when it swallows the shock, and in fact the two horizon radii are related by
\be
\tilde{r}_{h} = r_{h}\sqrt{\frac{M+E}{M}} = r_{h}\sqrt{1+\epsilon},
\ee
where we've defined $\epsilon \equiv E/M$.

We will parametrize the shock wave by its boundary time $t_w$ and its energy $\epsilon$. In the $(u,v)$ coordinates, we can describe the shock wave as traveling along a null surface of constant $u=u_0$. In the $(\tilde{u}, \tilde{v})$ coordinates, the shock wave travels on a null surface at constant $\tilde{u}=\tilde{u}_0$. We see from \eqref{uv} and \eqref{tilde} that by demanding that the shock leave the boundary at time $t_w$ that these surfaces are given by
\be
u_0 = e^{-r_{h} t_w / \lads^2}, \qquad \tilde{u}_0 = e^{-\tilde{r}_{h} t_w / \lads^2}, \label{u-shock-coordinate}
\ee
respectively.

Next, we need to relate the $(\tilde{u}, \tilde{v})$ coordinates to the $(u,v)$ coordinates. The matching conditions \cite{Dray:1985yt} amount to ensuring that the metric is continuous. First, we match across the radius of the transverse space. This gives
\be
\tilde{r}_{h}\, \frac{1 - \tilde{u}_0 \tilde{v} }{1+\tilde{u}_0 \tilde{v} } = r_{h}\, \frac{1 - u_0 v}{1+u_0 v} \label{matching}
\ee
Solving for $\tilde{v}$, we see that it's a fractional linear transformation of $v$,
\be
\tilde{v}(v) = \frac{1}{\tilde{u}_0}\bigg(\frac{(\tilde{r}_{h}-r_{h}) +  (\tilde{r}_{h}+r_{h})u_0 v}{   (\tilde{r}_{h}+r_{h}) + (\tilde{r}_{h}-r_{h})u_0v}\bigg) \label{vp_of_v}
\ee
Now normally we'd expand \eqref{vp_of_v} in the limit $\epsilon\to0$ and $t_w\to \infty$, with $\epsilon / u_0$ fixed. Then, we'd find $\tilde{v}(v) = v + \epsilon / 4u_0$ and that crossing the shock induces a simple shift. At this order, we'd also have $\tilde{r}_{h} = r_{h}$, and  we'd  pick $\tilde{u}(u) = u$ to satisfy the final matching condition. However, since $t_w\to \infty$, the shock would have to lie on the horizon $u=0$. Since $\epsilon\to0$, the shock would have to have vanishing mass.

Instead, let's study finite energy shocks at finite time, keeping $\epsilon$ and $t_w$ finite and arbitrary. The only thing left to do is to relate $\tilde{u}$ and $u$. The final matching condition \cite{Dray:1985yt} is that $g_{uv}\, du dv = g_{\tilde{u}\tilde{v}} \, d\tilde{u}d\tilde{v}$ along the shock
\be
-\frac{4 \lads^2}{(1+uv)^2}dudv|_{u=u_0} = -\frac{4 \lads^2}{(1+\tilde{u}\tilde{v})^2}d\tilde{u}d\tilde{v}|_{\tilde{u}=\tilde{u}_0}
\ee
This reduces to the following condition on $\tilde{u}'(u)$
\be
\tilde{u}'(u)|_{u=u_0} = u_0^{\sqrt{1+\epsilon}-1}\sqrt{1+ \epsilon}. \label{u-prime-matching}
\ee
While there is still some freedom in the function $\tilde{u}(u)$ away from the shock, we can simply satisfy \eqref{u-prime-matching} by taking
\be
\tilde{u} = u^{\sqrt{1+\epsilon}}. \label{up_of_u}
\ee
This condition makes the first derivative smooth across the shock on the boundary, but causes a kink across the shock on the singularity.

Now that we have $\tilde{u}(u)$ and $\tilde{v}(v)$, we can find out how far the horizon jumps out after absorbing the shock by inverting $\tilde{v}(v) = 0$. This is the analog of the simple shift we are used to
\be
v(\tilde{v}=0) = -\frac{1}{u_0}\bigg( \frac{\sqrt{1+\epsilon}- 1}{\sqrt{1+ \epsilon} +1} \bigg), \label{tilde-shift}
\ee
which goes like $- \epsilon / 4 u_0$ in the limit $\epsilon \to 0$, exactly as expected from \cite{shock}.
The Kruskal diagram is shown in  Fig.~\ref{finite}. Below the shock in the white region, we use the $(u,v)$ coordinates. Above the shock in the gray region, we use coordinates $(\tilde{u}(u), \tilde{v}(v))$.   We see that (as expected) there's no kink across the shock at the boundary, but there is a kink at the matching across the singularity.

Let us make a brief remark on the action of $\CW$ in the geometry shown in Fig.~\ref{finite}. As we will explain in the next subsection, we find something very different from \eqref{eq:complexity-finite-time-shock}. Here, for times $t_L$ before the perturbation $t_w$ the rate of growth with $t_L$ would be $M$, but for times after the perturbation, the rate would be $M+E$. This is indicative of the fact that this geometry actually represents energy $E$ being injected into the state on the left boundary at time $t_L$.

\subsection{Precursor at finite time and energy}
While the geometry Fig.~\ref{finite} considered in the last section is a proper solution to Einstein's equations, it is not the correct dual to the state
\be
|\psi \rangle = W(t_w) \TFD, \label{precursor-state}
\ee
where $W(t_w) = e^{-iH_L t_w} \, W \, e^{iH_L t_w}$ and $W$ is an operator smeared over the spatial sphere with energy $\epsilon M$, and $\epsilon$ is fixed but not necessarily small. This is no longer a small perturbation---its energy scales with $N^2$---but we can still consider its backreaction and the geometry dual to the state.

The correct procedure for constructing the geometry is described in \cite{Shenker:2013yza}. First, we start with the thermofield double state, which is dual to the eternal AdS black hole geometry. Then, we evolve back by time $t_w$ and apply the operator $W$. Finally, we evolve forward by $t_w$ back to $t_L=0$. We can think of this in terms of a time fold, where the perturbation is added on the second sheet. The final geometry is given by continuing the final sheet before and after the perturbation. The upshot is that the perturbation travels away from the boundary into the future \emph{and} into the past.

 \begin{figure}
\begin{center}
\includegraphics[scale=.61]{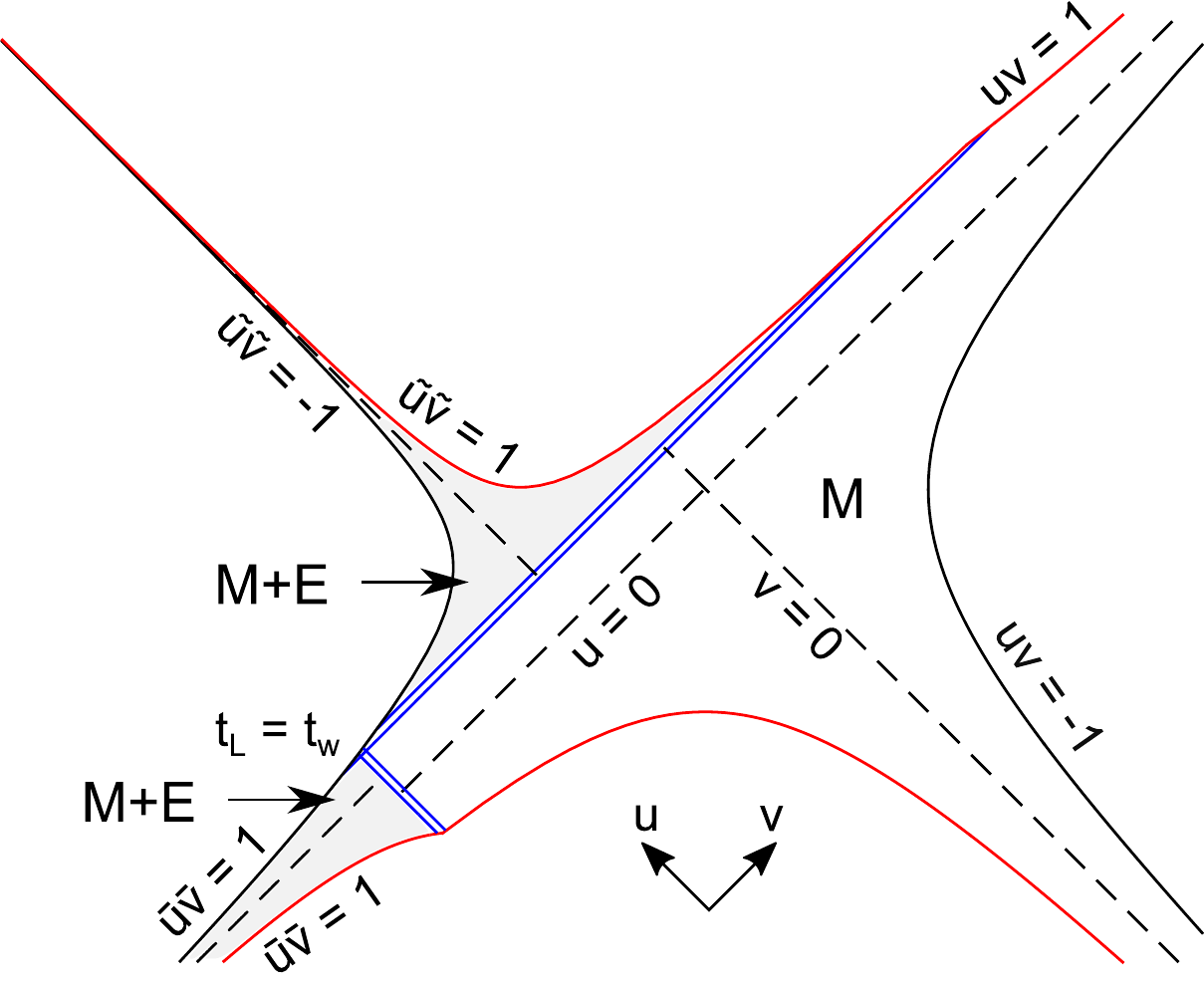}
\caption{Kruskal diagram of shock wave at finite time $t_w$. The white region to the right of the shock has energy $M$ and is covered by coordinates $(u,v)$. The gray region to the left of the shock has energy $M+E$. Above the shock, it's covered by coordinates $(\tilde{u}, \tilde{v})$. Below it's covered by $(\bar{u}, \bar{v})$. The shock is drawn in blue. When leaving the boundary it travels along the null surface $u_0 = e^{-r_{h} t_w / \lads^2}$ or equivalently $\tilde{u}_0 = e^{-\tilde{r}_{h} t_w / \lads^2}$. When approaching the boundary it travels along the null surface $v_0 = - e^{r_{h} t_w / \lads^2}$ or equivalently $\bar{v}_0 = - e^{\tilde{r}_{h} t_w / \lads^2}$. }\label{finite-from-singularity}
\end{center}
\end{figure}

The Kruskal diagram of this geometry is shown in Fig.~\ref{finite-from-singularity}. The correct picture for the state \eqref{precursor-state} is that of the perturbation emerging from the past singularity, materializing on the boundary at time $t_L=t_w$, and then traveling off into the future singularity. Additionally, note that the past horizon shrinks by some amount after ejecting the perturbation, and the future horizon grows outward by a different amount after swallowing the perturbation. These are the analogs of the simple shifts usually considered in the infinite time shock wave geometries.

The right way to think about the difference between the geometries in Fig.~\ref{finite} and Fig.~\ref{finite-from-singularity} is that in the former case, energy $E$ is injected into the left CFT at time $t_w$. Before that time, there the CFT had average energy $M$ and there was no perturbation. In the latter case, the state in the Schr\"{o}dinger picture is given by \eqref{precursor-state}. We can time evolve the state on the left side via the time evolution operator $e^{iHt_L}$, but at all times the perturbation is present and the system has energy $M+E$. This is important, because these two geometries give different predictions for the complexity of the state for times $t_L$ before $t_w$.

The coordinates above and to the left of the shock are still the $(\tilde{u}, \tilde{v})$ coordinates considered in the last section. The transformation between these coordinates and the $(u,v)$ coordinates is given by \eqref{vp_of_v} and \eqref{up_of_u}. Let's label the coordinates below the shock $(\bar{u}, \bar{v})$. The matching procedure is analogous to what we did before, so we will briefly state the results. The past moving shock travels along the surface
\be
v_0 = - e^{r_{h} t_w / \lads^2}, \qquad \bar{v}_0 = - e^{\tilde{r}_{h} t_w / \lads^2}.
\ee
The shock hits the boundary at $(u_0, v_0)$ and $(\bar{u}_0, \bar{v}_0)$. Using the definition of the boundary and direct comparison to \eqref{u-shock-coordinate}, we can easily identify the relations
\be
v_0 = -u_0^{-1}, \qquad \bar{v}_0 = - \tilde{u}_0^{-1}.
\ee
From this, we can simply write down the full coordinate transformations
\be
\bar{u}(u) = -\tilde{u}_0 \bigg( \frac{ (\tilde{r}_{h} - r_{h}) u_0 - (\tilde{r}_{h} + r_{h})u}{(\tilde{r}_{h} + r_{h}) u_0 - (\tilde{r}_{h} -r_{h})u } \bigg), \qquad \bar{v}(v) = -(-v)^{\sqrt{1+\epsilon} }.
\ee
Similar to before, there was freedom in the metric matching condition in how we extend the $\bar{v}(v)$ function away from the shock.

Finally, as noted in \eqref{tilde-shift}, the analog of the simple shift for the future horizon is given by $v(\tilde{v} = 0)$. For the past horizon, the shift is given by
\be
u(\bar{u} = 0) = \frac{u_0}{\epsilon} \big(1 - \sqrt{1+\epsilon}\big)^2,
\ee
which goes like $u_0 \epsilon / 4 \sim 0$ in the limit of $\epsilon  / 4u_0$ fixed.

\section{The past wedge} \label{sec:pastwedge}

\begin{figure}[h] 
   \centering
   \includegraphics[width=3.5in]{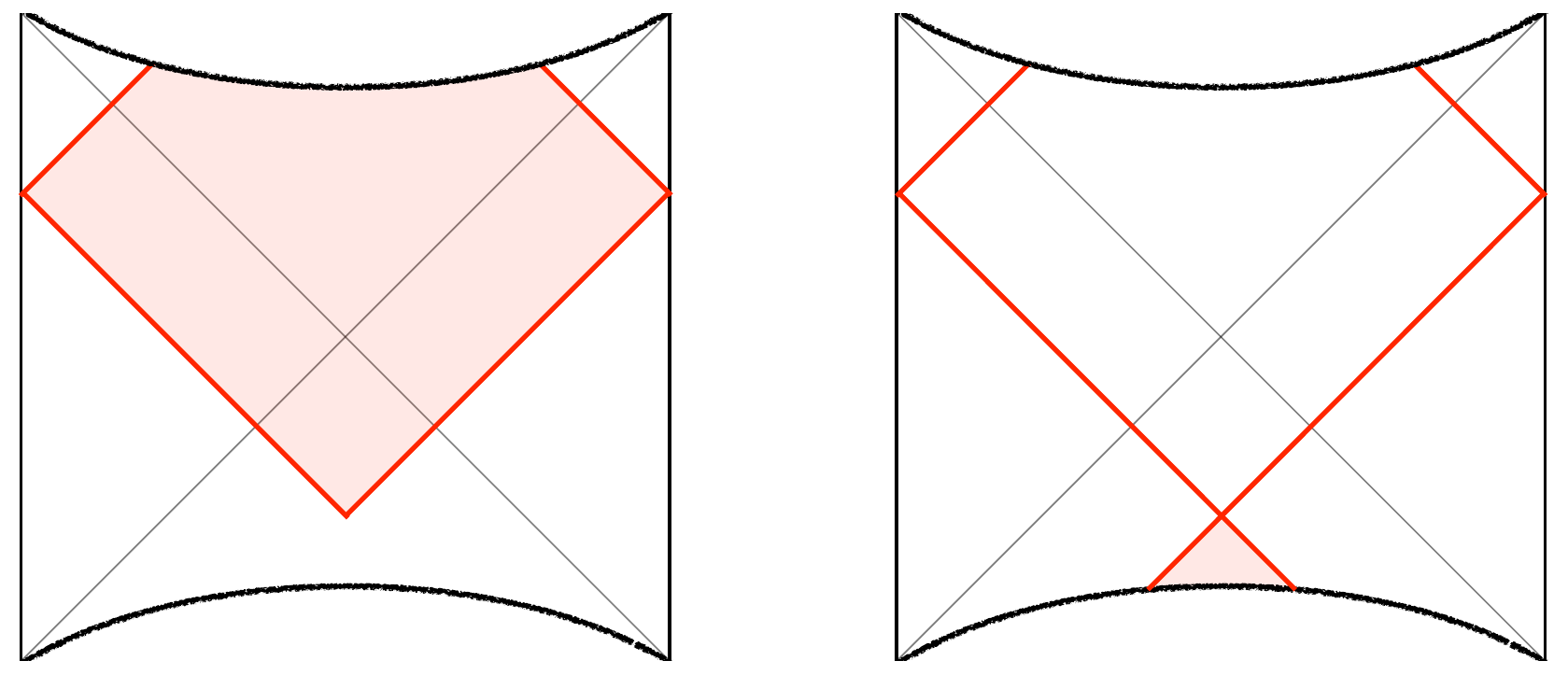}
      \caption{{\bf Left}: the  Wheeler-DeWitt patch (shaded). {\bf Right}: the past wedge (shaded) has the same rate of growth of action as the WDW patch and provides another way to think about where the computation is done.}
      \label{fig-ExcludedWedge}
\end{figure}

The rate of change of action of the Wheeler-DeWitt patch is  equal to the rate of change of action of the past wedge. The ``past wedge'' is  defined in Fig.~\ref{fig-ExcludedWedge}.
This provides both a calculational tool and an alternative perspective on our prescription.

Consider the two shaded patches in Fig.~\ref{fig-ExcludedWedge}. The patch on the left is the standard Wheeler-DeWitt patch. The patch on the right is that part of the bulk that can send a signal to both the left and right boundary observers. The rates of change of action of the two patches are identical. The rates are identical not just at late times but at all times. The Wheeler-DeWitt patch action and the past wedge action differ by a time-independent constant.

Here's how to see they are equal. The action outside the left light ``cone'' is constant (by left time translation symmetry); the action outside the right light ``cone'' is constant (by right time translation symmetry). The action outside both light `cones' (i.e. the Wheeler-DeWitt patch) is nonconstant only insofar as there is a patch that is outside both left and right light cones (the past wedge) and that is therefore double counted.

\bibliographystyle{ut}
\bibliography{complexity_long-version}

\end{document}